\documentclass[%
 reprint,
 aip,
 graphicx,
 amsmath,amssymb,
]{revtex4-2}

\usepackage{graphicx}% Include figure files
\usepackage{dcolumn}% Align table columns on decimal point
\usepackage{bm}% bold math

\usepackage{color}
\usepackage[flushleft]{threeparttable}

\usepackage[hidelinks]{hyperref}
\usepackage[capitalise]{cleveref}
\usepackage[utf8]{inputenc}
\usepackage[T1]{fontenc}
\usepackage{mathptmx}
\usepackage{textcomp}
\usepackage{etoolbox}
\makeatletter
\def\@email#1#2{%
 \endgroup
 \patchcmd{\titleblock@produce}
  {\frontmatter@RRAPformat}
  {\frontmatter@RRAPformat{\produce@RRAP{*#1\href{mailto:#2}{#2}}}\frontmatter@RRAPformat}
  {}{}
}%
\makeatother

\usepackage{soul}
\definecolor{lightgray1}{gray}{0.95}

\newcommand{\citeinline}[1]{Ref. \onlinecite{#1}}

\usepackage[most]{tcolorbox}

\tcbset {
  base/.style={
    arc=0mm, 
    bottomtitle=0.5mm,
    boxrule=0mm,
    colbacktitle=black!10!white, 
    coltitle=black, 
    fonttitle=\bfseries, 
    left=2.5mm,
    leftrule=1mm,
    right=3.5mm,
    title={#1},
    toptitle=0.75mm, 
  }
}

\definecolor{brandblue}{rgb}{0.34, 0.7, 1}
\newtcolorbox{mainbox}[1]{
  colframe=brandblue, 
  base={#1}
}

\newtcolorbox{subbox}[1]{
  colframe=black!30!white,
  base={#1}
}
 
\begin{document}

\preprint{AIP/123-QED}
\title{Feedback Cooling of an Insulating High-Q Diamagnetically Levitated Plate}

\author{S. Tian}
\affiliation{ 
Quantum Machines Unit, Okinawa Institute of Science and Technology Graduate University, Onna, Okinawa 904-0495, Japan 
}%

\author{K. Jadeja}
\affiliation{ 
Quantum Machines Unit, Okinawa Institute of Science and Technology Graduate University, Onna, Okinawa 904-0495, Japan 
}%

\author{D. Kim}
\affiliation{ 
Quantum Machines Unit, Okinawa Institute of Science and Technology Graduate University, Onna, Okinawa 904-0495, Japan 
}%

\author{A. Hodges}
\affiliation{ 
Quantum Machines Unit, Okinawa Institute of Science and Technology Graduate University, Onna, Okinawa 904-0495, Japan 
}%

\author{G. C. Hermosa}
\affiliation{Department of Chemical Engineering and Materials Science, Yuan Ze University, Chung-Li 32003, Taiwan }

\author{C. Cusicanqui}
\affiliation{Tecnologico de Monterrey, Escuela de Ingenieria y Ciencias, Monterrey 64849, Mexico}

\author{R. Lecamwasam}
\affiliation{A*STAR Quantum Innovation Centre (Q.InC), Institute for Materials Research and Engineering (IMRE), Agency for Science, Technology and Research (A*STAR), 2 Fusionopolis Way, 08-03 Innovis 138634, Republic of Singapore}
\affiliation{ 
Quantum Machines Unit, Okinawa Institute of Science and Technology Graduate University, Onna, Okinawa 904-0495, Japan 
}%

\author{J. E. Downes}
\affiliation{Department of Physics and Astronomy, Macquarie University, Sydney, NSW 2109, Australia}

\author{J. Twamley}
\affiliation{ 
Quantum Machines Unit, Okinawa Institute of Science and Technology Graduate University, Onna, Okinawa 904-0495, Japan 
}%

\begin{abstract}
    {\color{black} Levitated systems in vacuum have many potential applications ranging from new types of inertial and magnetic sensors through to fundamental issues in quantum science, the generation of massive $\rm  Schr\ddot o dinger$ cats, and the connections between gravity and quantum physics. In this work, we demonstrate the passive, diamagnetic levitation of a centimeter-sized massive oscillator which is fabricated using a novel method that ensures that the material, though highly diamagnetic, is an electrical insulator. By chemically coating a powder of microscopic graphite beads with silica and embedding the coated powder in high-vacuum compatible wax, we form a centimeter-sized thin square plate which magnetically levitates over a checkerboard magnet array. The insulating coating reduces eddy damping by almost an order of magnitude compared to uncoated graphite with the same particle size. These plates exhibit a different equilibrium orientation to pyrolytic graphite due to their isotropic magnetic susceptibility. We measure the motional quality factor to be $Q\sim 1.58\times 10^5$ for an approximately centimeter-sized composite resonator with a mean particle size of 12 microns.  
    Further, we apply delayed feedback to cool the vertical motion of frequency $\sim 19\,{\rm Hz}$ from room temperature to 320 millikelvin.
    }
\end{abstract}

\maketitle

Optomechanical systems are the most precise measuring devices in existence, most notably the LIGO gravitational wave observatory which can measure distortions in spacetime 10,000 smaller than a proton. Tabletop systems aim to exploit the versatile optomechanical interaction to design compact quantum sensors. Optomechanical systems are fundamentally limited by thermal noise from the environment, which interferes with the measurement signal and causes decoherence of fragile quantum states. Levitated optomechanics suspends the system without any physical connection to the environment. The resulting exquisite isolation means that the quantum ground state can be attained at room temperature using feedback-cooling.\cite{Delic2020,Magrini2021Real-timeTemperature} This is promising for both high-precision laboratory measurements, and deployment in noisy real-world settings.\cite{Gonzalez-Ballestero2021,Winstone2023LevitatedPerspective} Levitation also leads to a natural coupling with gravity. Thus levitated systems are natural testbeds for probing some of the deepest outstanding questions in physics, such as quantum gravity and collapse mechanisms.\cite{Moore2021}

Diamagnetic levitation of graphite slabs has been gaining increasing attention as a platform for levitated optomechanics. Most levitated platforms use active methods such as optical tweezers or Paul traps. These use time-varying optical or electromagnetic fields,\cite{Gonzalez-Ballestero2021,Winstone2023LevitatedPerspective} and can trap particles of nano- and micro-meter sizes. In contrast diamagnetic levitation is passive, and centimeter-sized slabs of graphite can be easily levitated above an array of commercially-available permanent magnets.\cite{Young2019OpticalGraphite,Fujimoto2019DiamagneticGraphite,Yee2021PhotothermalRevised,Ewall-Wice2019OptomechanicalGraphite,Chen2020} This removes the noise associated with an active power source.\cite{Simon2000} Moreover, the lowered power and hardware requirements are promising for developing commercial sensors. Diamagnetic levitation can also support much larger masses than traditional levitated optomechanics. More massive systems have greater sensitivity for accelerometry and gravimetry,\cite{Timberlake2019} and are crucial to explore the behaviour of quantum physics at larger scales. The large mass, ease of use, and passive nature give graphite diamagnetic levitation a unique status in levitated optomechanics.

The greatest limiting factor for these systems is eddy damping. 
Most experiments use Highly Oriented Pyrolytic Graphite (HOPG), a synthetic material consisting of layered planes of graphite. This has a very strong diamagnetic susceptibility along the axis perpendicular to the graphite planes. However, HOPG is an electrical conductor. As it moves through the magnetic field generated by the magnets, currents are induced along the graphite planes, which
causes strong damping.
Consequently the quality factor of levitated pyrolytic graphite in vacuum is only several hundred for millimeter-sized slabs, and decreases as size increases.\cite{Chen2020} Although a thin graphite film has demonstrated sufficient sensitivity to experimentally test theories of dark matter,\cite{Yin2022ExperimentsEnergy} increasing the quality factor would substantially increase sensitivity, and unlock powerful feedback-cooling methods.

It is thus necessary to find methods of suppressing eddy damping in levitated graphite systems. The currents can be disrupted by engineering narrow slits, allowing quality factors of several thousand in vacuum.\cite{Romagnoli2023ControllingPlate, Xie2023SuppressingGeometry} An alternative approach has been to use composite materials, consisting of micrometer-sized graphite particles dispersed in an insulating resin.\cite{Chen2022DiamagneticResonators} Eddy currents can only flow within the microparticles, or between neighbouring particles which happen to touch. Millimeter-sized slabs of this composite graphite attained quality factors of half a million at high vacuum at room temperature, the largest demonstrated for any system of that mass.\cite{Chen2022DiamagneticResonators}

Composite graphite is thus a highly promising system for levitated optomechanics. However, it also presents new challenges. Since the magnetic susceptibility of graphite is highly oriented, the random positioning of particles in the composite lowers the effective magnetic susceptibility. The susceptibility further decreases for particle sizes smaller than $30\:\mathrm{\mu m}$,\cite{Semenenko2018DiamagnetismRevised} though such sizes are necessary to significantly suppress eddy currents. Thus the mass that can be supported is decreased, if for example we wished to place a mirror or other system on the levitated graphite. There is also a limit on the volume fraction of graphite in the material, which was about $40\%$ in Chen et al\cite{Chen2022DiamagneticResonators} for reasons of structural integrity. Moreover, the graphite particles must be kept separated by the insulating resin. At high volume fractions it becomes increasingly likely that neighbouring graphite particles will touch, allowing eddy currents to flow between them and thus decreasing the quality factor.

\begin{figure}
    \begin{center}	
        \includegraphics[width=\columnwidth]{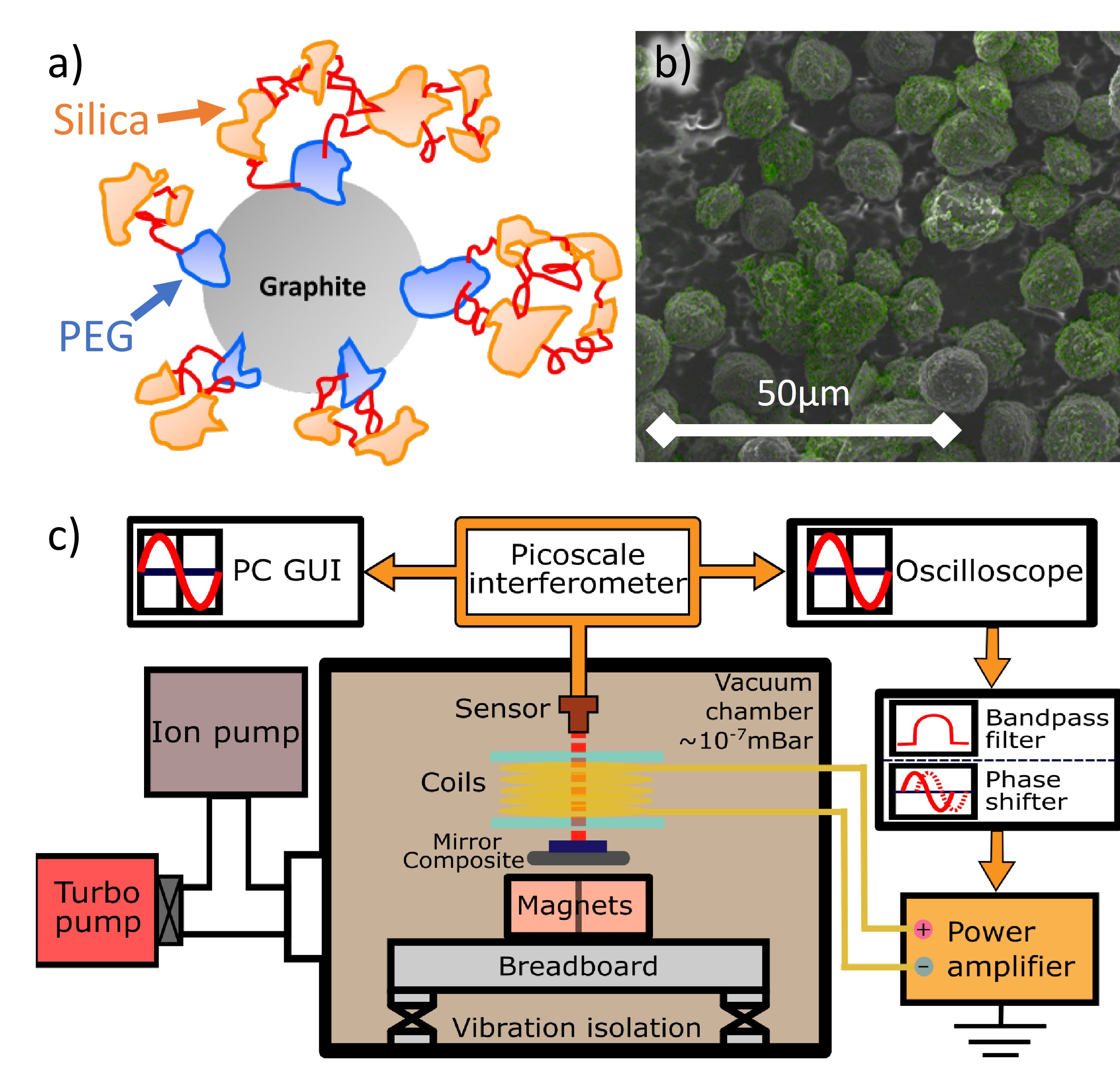}
    \end{center}
    
    \vspace*{-10pt}
    
    \caption{Coated graphite and experimental setup.
    a) We chemically coat each graphite particle with a layer of electrically insulating silica. 
    Polyethylene glycol (PEG) works as a bridgemer, allowing silica to bind to the surface of the graphite. The coated graphite particles are mixed with vacuum-compatible wax and shaped into insulating diamagnetic slabs (see Supplementary for details). 
    b) Scanning electron microscope image of the coated graphite microbeads, overlaid with Energy-Dispersive X-ray (EDX) elemental mapping. Green regions indicate silicon, confirming the presence of the insulating coating. 
    c) The insulating diamagnetic slab levitates above a checkerboard of four NdFeB magnets with orientation alternating between the north and south poles. The system is placed on a vibration-isolation platform and then kept in high vacuum ($10^{-6}-10^{-7}\:\mathrm{mbar}$), see \citeinline{Romagnoli2023ControllingPlate} for more details of this setup. A mirror is fixed onto the slab for interferometric real-time measurement of the position and velocity. The delayed velocity signal is fed back after filtering and time delay through a coil of wire, which applies magnetic actuation to cool the vertical motion of the slab.
    }\label{fig:setup}
        
\end{figure}

To enable large graphite volume fractions while maintaining suppression of eddy currents, we coat the graphite particles with an insulating shell. Our graphite particles are mesocarbon microbeads (MCMB), whose average diameter is $11\pm2\:\mathrm{\mu m}$. We chemically coat these with a thin insulating layer of silica in a `sol-gel process' using polyethylene glycol (PEG) and the silica precursor tetraethyl orthosilicate (TEOS),\cite{Kim2017AGraphite} see the supplementary material for details on the fabrication process. First PEG adsorbs onto the surface of the microcarbon microbeads. A solution of TEOS and ammonium hydroxide catalyst is then added, and stirred for 17 hours on a hot plate. During this time silicon from TEOS attaches to the PEG, resulting in a silica coating on the graphite as shown in \cref{fig:setup} a). The mixture is then washed, filtered, and dried. Analysis using scanning electron microscopy with elemental mapping shown in \cref{fig:setup} b) confirms the near-uniform coverage of the graphite beads with silica. The powder is then mixed in vacuum-compatible wax at a temperature of approximately $150 ^\circ\,{\rm C}$, then cooled and shaped into small square slabs of approximate width $8\:\mathrm{mm}$ and thickness $0.5\:\mathrm{mm}$. The mass fraction of graphite is $60\%$, with the volume fraction estimated as $43.3\%$. At fractions larger than this mixing the graphite powder into the wax requires significant stirring force, which causes cracking of the silica shell.

\begin{figure}
    \begin{center}
    \includegraphics[width=\columnwidth]{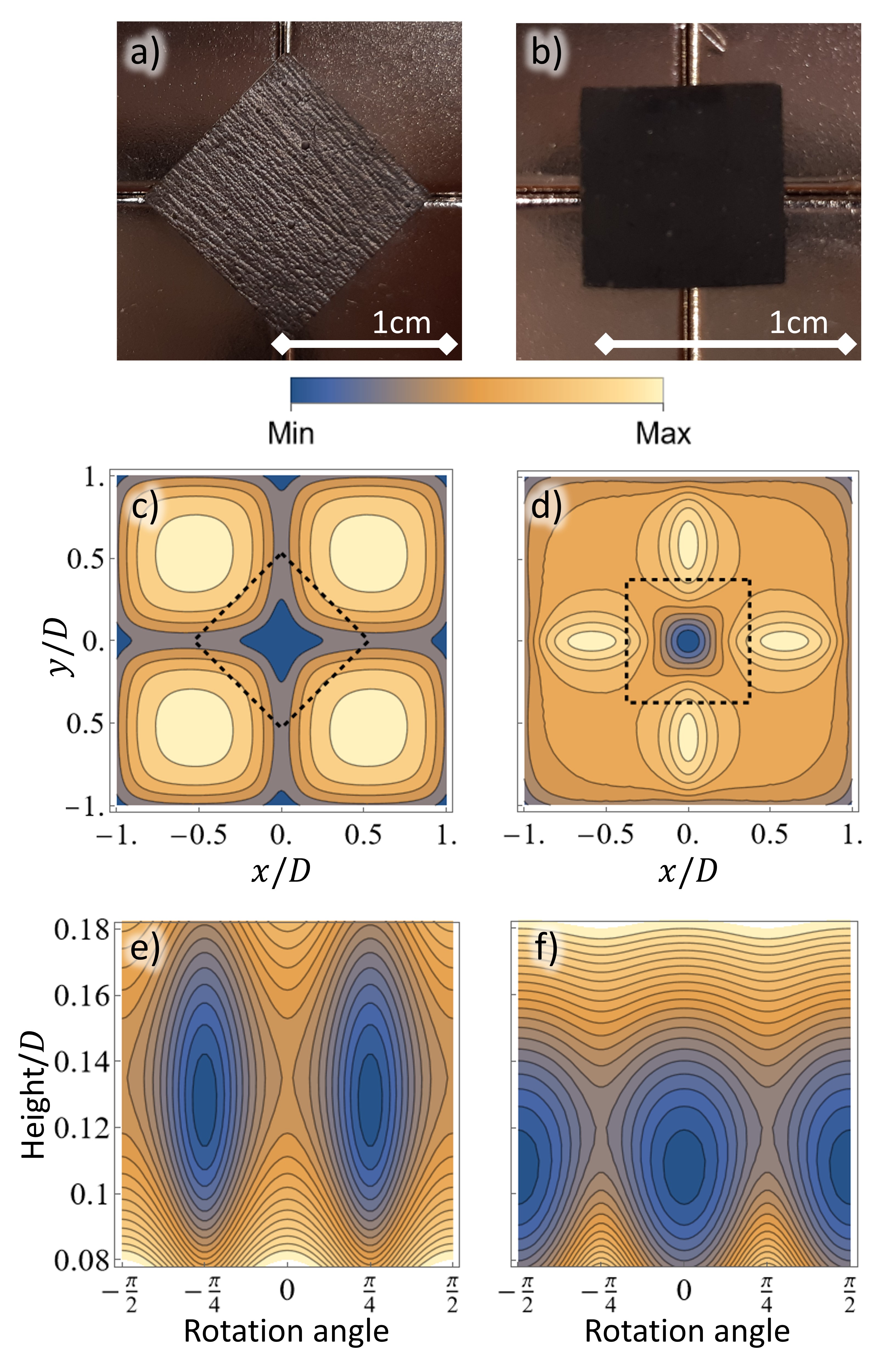}
    \end{center}
    \vspace{-20pt}
    \caption{Re-orientation of diamagnetic slabs
    a) Photograph of levitated HOPG with dimensions $12.4\times 12.4\times 0.7\:\mathrm{mm}^3$, levitated above magnets with side lengths $D=12.7\:\mathrm{mm}$. At equilibrium HOPG is oriented diagonally with respect to the magnets. 
    b) Photograph of levitated composite graphite with dimensions $8.5\times 8.6\times 0.6\:\mathrm{mm}^3$. The composite orients in line with the magnet array. 
    c) Contours of the magnetic potential energy density \cref{eq:MagneticPotentialEnergy} for HOPG. Due to the oriented magnetic susceptiblity, the $z$-component of the magnetic field is weighted five times more strongly than the $x$- and $y$-components. The dashed lines represent a graphite slab with side-lengths $0.75D$, which minimises the magnetic potential energy when oriented diagonally. 
    d) Magnetic potential energy density for composite graphite. The magnetic susceptilibity weights all components of the magnetic field evenly. Now the rotation minimising total magnetic energy lines up with the magnets.
    e) Potential energy of the HOPG as a function of orientation and height, for a slab of side length $0.75D$. We use the experimentally measured density of $2070\:\mathrm{kg/m^3}$. Minima occur when the slab is rotated at $\pi/4$ relative to the magnets, matching our observations. 
    f) Potential energy of composite graphite, whose density was measured to be $1442\:\mathrm{kg/m^3}$, and mass fraction of $60\%$ graphite in the composite. The energy is minimised when the slab is in line with the magnets. The composite levitates lower than HOPG, and the angular confinement is much weaker.
    Note that c)-f) are each normalised, so the colour scale cannot be directly compared between plots.}
    \label{fig:orientation}

\end{figure}

The graphite slabs are levitated above a checkerboard-array of four permanent magnets, whose upper faces alternate between north and south poles. In \cref{fig:orientation} a) and b) we show photographs of both pyrolytic and composite graphite, taken from above once their position had reached equilibrium. 
We can see that the pyrolytic and composite graphite exhibit different orientations.
This arises from the difference in their magnetic susceptibilities. The potential energy of a diamagnet in a magnetic field is given by 
\begin{equation}\label{eq:MagneticPotentialEnergy}
    U_B(z,\phi) = -\frac{1}{2\mu_0}\int_{\mathcal{V}(z,\phi)}\left[\chi_{x}B_x^2+\chi_{y}B_y^2+\chi_zB_z^2\right]\,\mathrm{d}x\,\mathrm{d}y\,\mathrm{d}z.
\end{equation}
Here $\chi_j$ are the components of the magnetic susceptibility tensor, $B_j$ the components of the magnetic field, and $\mu_0$ is the vacuum permeability. The integral is over the volume $\mathcal{V}$ of the slab, which depends on its levitation height $z$ and rotation $\phi$. The susceptibility of Pyrolytic graphite is highly oriented: $\chi_j^{\mathrm{pyro}}=-(85,85,450)\times 10^{-6}$, and thus it orients itself primarily to avoid the $z$-component of the magnetic field. The composite graphite on the other hand has uniform susceptibility\cite{Chen2022DiamagneticResonators} $\chi_j^{\mathrm{comp}}=-(120,120,120)\times 10^{-6}$, and orients itself to avoid all magnetic field components equally. We plot the magnetic potential energy densities in \cref{fig:orientation} c) and d), which we see accounts for the difference in orientation. 

We further study the difference between the two materials by evaluating the total potential energy, which is the sum of the magnetic and gravitational components. See the Supplementary for details of this calculation. We graph this for pyrolytic and composite graphite in \cref{fig:orientation} e) and f), which clearly show the difference in orientation. We can see that the composite graphite experiences a much looser angular confinement. The composite also levitates lower. There are several effects which contribute to the different levitation height, namely the lower effective susceptibility of the composite, its graphite mass fraction of 60\%, and the density of the wax. These factors also influence the shape of the trap, as we discuss in the Supplementary material.

We note that these observations seem to disagree with those of Chen \textit{et al.} 2022,\cite{Chen2022DiamagneticResonators} which reported composite resonators to orient themselves similarly to pyrolytic graphite. A more detailed analysis in our Supplementary material does not find a regime where such orientation is expected. One likely explanation is the geometry of the magnets. In our experiment shown in \cref{fig:orientation} a-b) the chamfering is small compared to the size of the magnets and graphite slabs, hence we assumed the magnets to be perfect cubes in our analysis. However, in Chen \textit{et. al} Figure 1d  the magnets show significant chamfering, comparable to the size of their composite slabs. Given how shallow the composite potential is in \cref{fig:orientation} f), it is likely that this could be distorted by such changes to the magnet geometry. Another explanation could be that some unknown factor is causing slight orientation of the particles in their composite.

\begin{figure}
    \begin{center}	
    \includegraphics[width=\columnwidth]{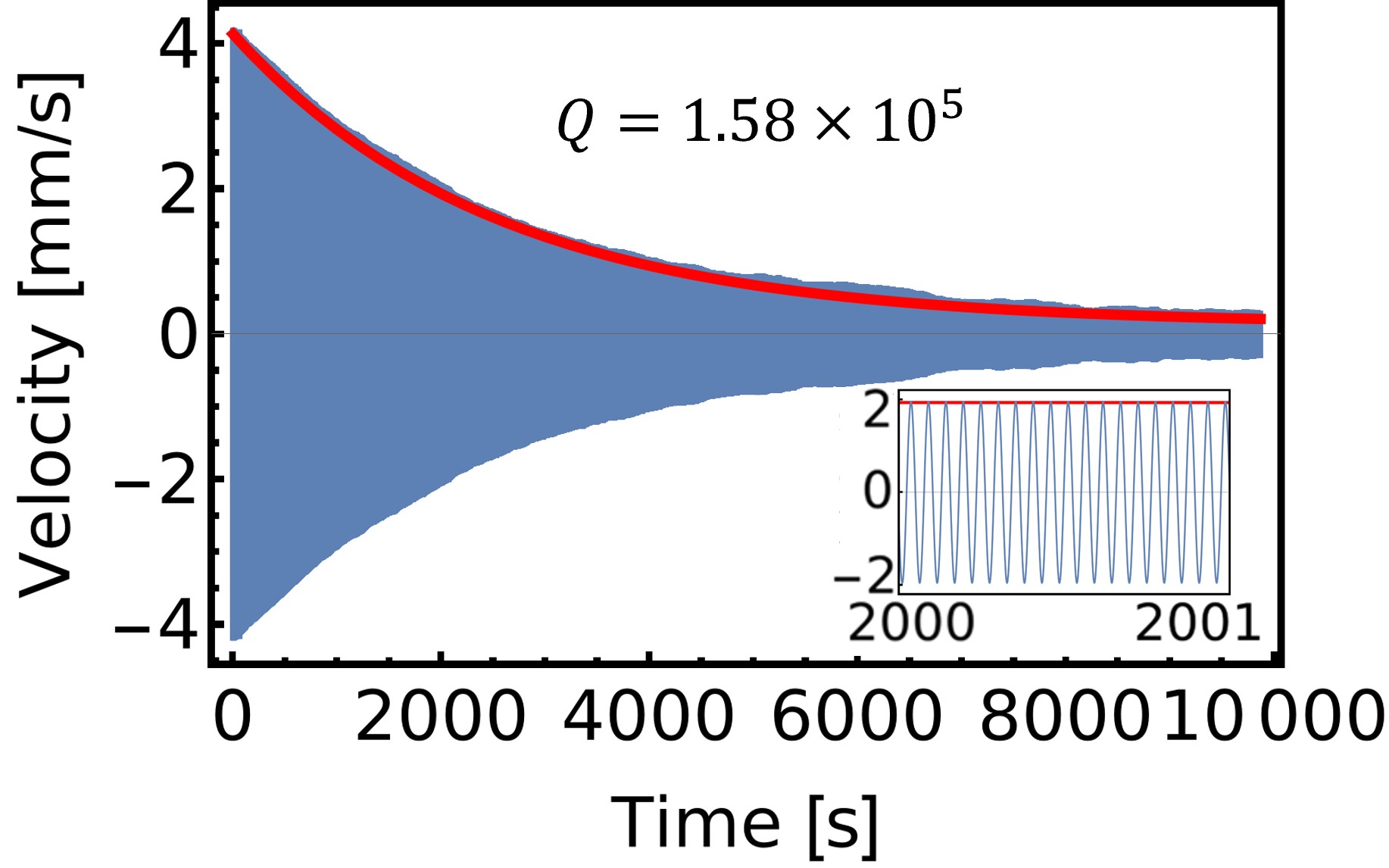}
    \end{center}
\vspace*{-10pt}
\caption{Ringdown of the graphite resonator at a pressure of $1.2\times 10^{-6} \:{\rm mbar}$. 
The blue line shows velocity as measured by the interferometer. The inset zooms-in over a span of $1 \:{\rm s}$. The red curve denotes the fitted envelope, from which we find a damping rate of $\gamma\sim 3.7\times 10^{-4} \:\rm Hz$, corresponding to a  quality factor of $Q\sim 1.58\times 10^{5}$. This is almost an order of magnitude smaller than the damping rate for other graphite composites with similar particle sizes ($11\pm 2\:\mathrm{\mu m}$).\cite{Chen2022DiamagneticResonators} 
This demonstrates that the insulating silica shell is effectively suppressing eddy currents between neighbouring graphite particles.
}\label{Ringdown}

\end{figure}

\textcolor{black}{The composite's trapped dynamics was characterised using the setup shown in \cref{fig:setup} c).}
The graphite levitates above a checkerboard of four permanent magnets, with a small mirror placed on top of the slab in order to read out its position and velocity with a Picoscale interferometer. 
This setup is most sensitive to the slab's vertical oscillation, which has a frequency of $18.95\:\mathrm{Hz}$. We actuate the system using a coil situated just above the graphite. A current flowing through this coil generates a magnetic field along its vertical axis, which perturbs the field generated by the magnets and thus exerts a force on the graphite.
As discussed in the supplementary material, the force exerted by the coil depends only on the magnitude of the applied voltage, not its sign. When applying voltage signals to the coil, we thus operate around a DC shift, allowing us to apply both positive and negative forces. When performing measurement at high vacuum only the ion pump is left on, allowing us to maintain vacuum while minimising vibrations. We refer to the supplementary material for further details on the setup.

To study the eddy damping in our novel composite, we drive the vertical mode using the coil at the resonance frequency of $\sim 18.9\:\mathrm{Hz}$, and then measure the ringdown at a pressure of $1.2\times 10^{-6} \:{\rm mbar}$ for $10,000 \rm s$, as shown in \cref{Ringdown}. The velocity of a damped resonator decays exponentially as $ v(t) \propto e^{-\gamma t}$, where $\gamma$ is the decay rate. By fitting the envelope of measured velocity, we find $\gamma \sim 3.7\times 10^{-4} \:\rm Hz$. The quality factor of the vertical mode is defined as $Q=\pi f_0/\gamma$, where $f_0$ is the natural frequency of the vertical mode. We find $Q\sim 1.58\times 10^{5}$ for the $8\:\mathrm{mm}$ square composite plate.

Our measured damping rate is three orders of magnitude smaller than HOPG with engineered slots,\cite{Romagnoli2023ControllingPlate,Xie2023SuppressingGeometry} showing the effectiveness of composite materials in suppressing eddy currents.
Moreover, the damping rate is almost one order of magnitude lower than that of the composite with the same particle size measured in \citeinline{Chen2022DiamagneticResonators} (see the Supplementary material for a detailed comparison). This demonstrates that the insulating silica shell is indeed suppressing the flow of eddy currents between neighbouring graphite particles, leading to almost an order of magnitude increase in quality factor.

\begin{figure}
\vspace{-10pt}
\includegraphics[width=\columnwidth]{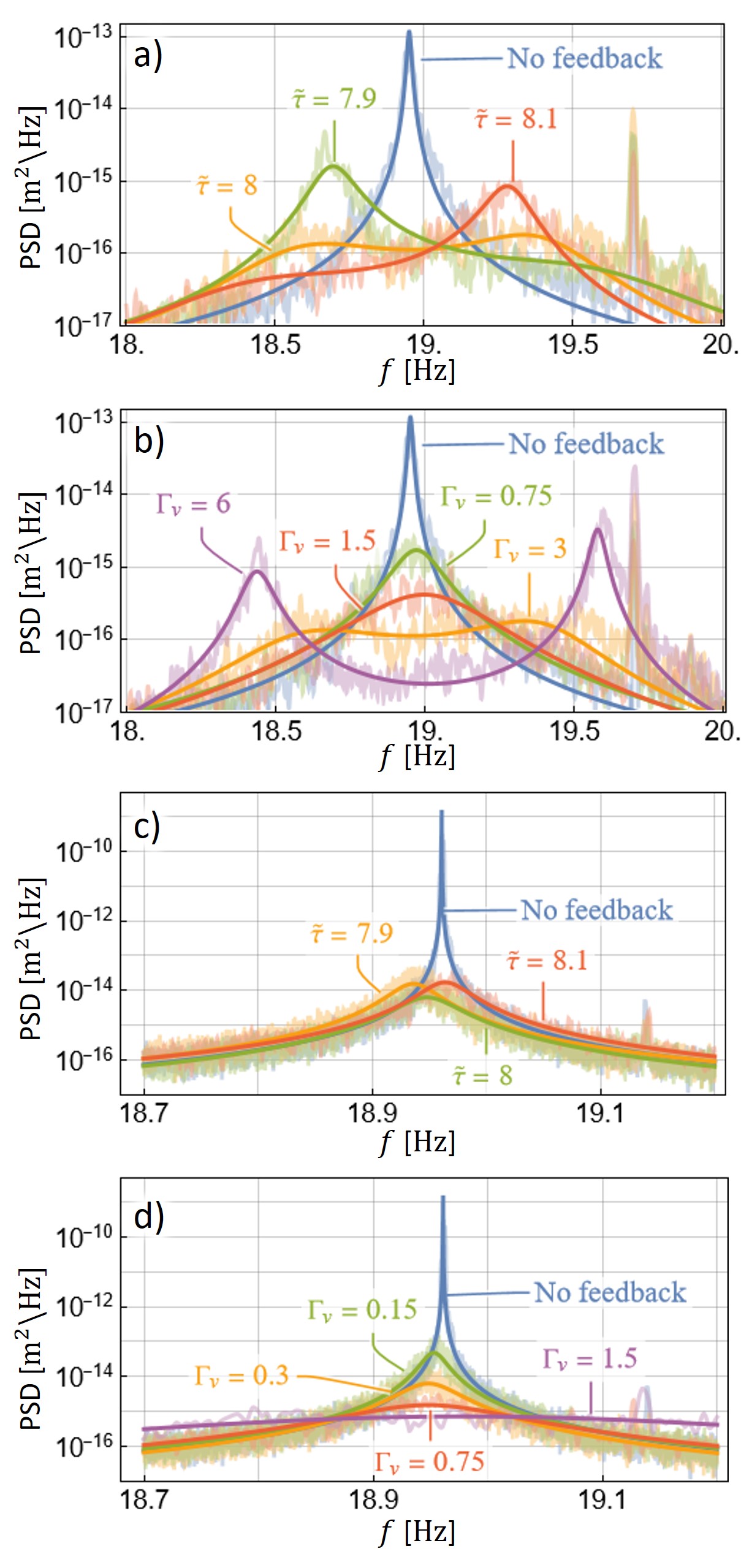}
\vspace*{-20pt}
\caption{Power spectral densities at a)-b) moderate pressure $10^{-2}\:\mathrm{mbar}$ and c)-d) low pressure $10^{-6}\:\mathrm{mbar}$. \textcolor{black}{We parameterise these with the feedback strength $\Gamma_v$ [Hz], and dimensionless time delay $\tilde{\tau}=\tau f_0$, where $f_0$ is the natural frequency [Hz].} The noisy traces are the experimental data, smoothed using Welch's method. The curves are fittings to \cref{eq:PSD}, which show good agreement with theory. In a) and c) we fix $\Gamma_{v}=(3,\:0.75)\:\rm Hz$ respectively, and vary the time delay. In b) and d) we fix $\tilde{\tau}=8$, and vary the feedback strength. We note the presence at moderate pressure of other modes.
Sidelobes appear due to the significant time delay in the system. At low pressure the peak is significantly narrowed, and feedback cooling is able to attain a temperature of $320\:\mathrm{mK}$. All fitting parameters are provided in the supplementary material.
}\label{fig:PSD_onecolumn}
\end{figure}

Optomechanical applications of graphite composites will invariably require feedback cooling \textcolor{black}{and aided by the electromagnetic coil above the levitated resonator we apply feedback forces to cool the vertical motion.} The feedback signal is generated from the \textcolor{black}{real-time} measured velocity by a Red Pitaya FPGA. To suppress noise and isolate the vertical mode, the velocity is bandpass filtered using a finite impulse response (FIR) filter. The filter and electronics have an intrinsic delay which we find to be approximately $7.6$ periods of the $18.9\:\mathrm{Hz}$ oscillation signal. The FPGA thus also applies a variable time delay, allowing us to make sure velocity feedback is applied with the correct phase relative to the composite's motion. The signal from the FPGA is then amplified, and applied to the coils. Further details on the feedback scheme are provided in the supplementary material.

For theoretical analysis, we approximate the vertical motion as a one-dimensional harmonic oscillator with a thermal drive. This approximation should be valid for small oscillations, where coupling to other motional modes should not be significant.
Considering the velocity feedback, the equation of motion is
\begin{equation}\label{eq:EOM}
        \ddot{x}(t)+\gamma\,\dot{x}(t)+\omega_0^2x(t)+\Gamma_v\dot{x}(t-\tau) =\sqrt{\frac{2\gamma k_BT}{m}}\xi(t),
\end{equation}
where dots denote the time derivative. The vertical position of the oscillator is given by $x$. The oscillator has resonant frequency $\omega_0=2\pi f_0$, damping rate $\gamma$, and mass $m$. The temperature of the thermal bath is $T$, and $k_B$ is Boltzmann's constant. The bath is modeled as Gaussian white noise, denoted $\xi(t)$. The feedback strength is given by $\Gamma_v$, which is applied with some time delay $\tau$. This corresponds to a feedback force of $m\Gamma_v \dot{x}(t-\tau)$. In the supplementary material we show that this has power spectral density:
\begin{equation}\label{eq:PSD}
        S_{xx} (\omega)= \frac
        {2k_BT\gamma/m}
        {\left[\omega_0^2-\omega^2+\omega\Gamma_v\sin(\omega\tau)\right]^2+
        \left[\omega\gamma+\omega\Gamma_{\nu}\cos(\omega \tau)\right]^2}.
\end{equation}
As discussed in the supplementary material this formula is valid when $n-1/4\lesssim\tau f_0\lesssim n+1/4$, where $f_0$ is the oscillator frequency and $n$ is an integer. In this case feedback opposes velocity. Otherwise the feedback causes unbounded heating of the system.

To characterise the effectiveness of feedback in this system, we performed delayed velocity feedback cooling experiments at moderate ($\sim 1\times 10^{-2}\: \rm mbar$) and low ($\sim 1\times 10^{-6}\: \rm mbar$) pressures, with varying time delays and feedback strengths. 
The measured power spectral densities are shown in \cref{fig:PSD_onecolumn}, which match closely with \cref{eq:PSD}.  
We describe the feedback using feedback strength $\Gamma_v\:  \rm[Hz]$, and dimensionless time delay $\tilde{\tau}=\tau f_0$, where $f_0$ is the natural frequency in the magnetic trap.
The value of $\gamma$ for low pressure is closely restricted to the damping from the ringdown measurement, while the value at moderate pressure is freely fit to the PSD. The parameter $\Gamma_v$ is fit from the experimental data. For the time delay $\tau$, we first find an approximate value using our measured delay and the manual delay added by the FPGA, and then fit $\tau$ within one period deviation of this. It is also necessary to \textcolor{black}{fit the overall scale of the PSD}.  %apply a constant scaling to the mass. 
\textcolor{black}{This scale fitting} accounts for the fact that \cref{eq:EOM} models a point particle following ideal Brownian motion, whereas our system consists of a three-dimensional \textcolor{black}{extended} plate. Measurements at the moderate pressure took 20 minutes, while at low pressure we required 7 hours to resolve the peak. As we show in the supplementary material, at low pressure without feedback, the resonance frequency drifts slightly during this time which may be caused by the thermal expansion of the composite plate.

The temperature is defined as being proportional to the integrated area under the PSD. We suppose that the system without feedback is in thermal equilibrium with the environment at $300 \: {\rm K}$. We find that cooling is strongest when $\tau$ is an integer number of periods, which from \cref{eq:EOM} means that the feedback is directly opposing the velocity. When $\tau$ deviates from an integer value the PSD broadens, and the peak shifts. At higher pressures with strong feedback strength, the time delay leads to visible sidelobes. Increasing the strength of the feedback leads to stronger cooling, and larger sidelobes at higher pressure. At low pressure, with a time-delay of $8$ periods and a feedback strength of $\Gamma_v=0.75 \:\rm Hz$, we attain a minimum temperature of $T\sim 320\:\mathrm{mK}$. See the supplementary material for further discussion of the experimental data and analysis.

In conclusion, we demonstrate a novel magnetically levitated centimeter-size composite resonator by mixing the silica-coated graphite particles with wax. The insulating coating on the graphite prevents eddy currents flowing between adjacent particles, significantly reducing eddy damping, and hence increasing the quality factor by an order of magnitude.
We report the cooling of vertical motional mode at different delays and feedback strengths, with good agreement between experiment and theory. At suitable feedback conditions, we realize strong cooling of the cm-sized resonator to a centre-of-mass temperature of $\sim 320\,{\rm mK}$.  
This work is a step forward towards the preparation of a macroscopic quantum spatial superposition, and shows the potential of our novel resonator for fundamental quantum physics and ultrahigh precision sensing.

We are grateful for the help and support provided by the Scientific Computing and Data Analysis Section, the Scientific Imaging Section  and the Engineering Section  at OIST.

%\bibliography{references}
%aipnum4-2.bst 2019-01-14 (MD) hand-edited version of apsrev4-1.bst

\end{document}

\preprint{AIP/123-QED}
\title{Supplementary Material for Feedback Cooling of an Insulating High-Q Diamagnetically Levitated Plate}
\author{S. Tian}
\affiliation{ 
Quantum Machines Unit, Okinawa Institute of Science and Technology Graduate University, Onna, Okinawa 904-0495, Japan 
}%

\author{K. Jadeja}
\affiliation{ 
Quantum Machines Unit, Okinawa Institute of Science and Technology Graduate University, Onna, Okinawa 904-0495, Japan 
}%

\author{D. Kim}
\affiliation{ 
Quantum Machines Unit, Okinawa Institute of Science and Technology Graduate University, Onna, Okinawa 904-0495, Japan 
}%

\author{A. Hodges}
\affiliation{ 
Quantum Machines Unit, Okinawa Institute of Science and Technology Graduate University, Onna, Okinawa 904-0495, Japan 
}%

\author{G. C. Hermosa}
\affiliation{Department of Chemical Engineering and Materials Science, Yuan Ze University, Chung-Li 32003, Taiwan }

\author{C. Cusicanqui}
\affiliation{Tecnologico de Monterrey, Escuela de Ingenieria y Ciencias, Monterrey 64849, Mexico}

\author{R. Lecamwasam}
\affiliation{A*STAR Quantum Innovation Centre (Q.InC), Institute for Materials Research and Engineering (IMRE), Agency for Science, Technology and Research (A*STAR), 2 Fusionopolis Way, 08-03 Innovis 138634, Republic of Singapore}
\affiliation{ 
Quantum Machines Unit, Okinawa Institute of Science and Technology Graduate University, Onna, Okinawa 904-0495, Japan 
}%

\author{J. E. Downes}
\affiliation{Department of Physics and Astronomy, Macquarie University, Sydney, NSW 2109, Australia}

\author{J. Twamley}
\affiliation{ 
Quantum Machines Unit, Okinawa Institute of Science and Technology Graduate University, Onna, Okinawa 904-0495, Japan 
}%
\maketitle

\section{Materials for Insulating Resonators}\label{Section:materials}
\graphicspath{ {./Figures/} }
\subsection{Fabrication of the Insulating Resonators}
We design and fabricate a novel type of high-Q levitated mechanical resonator which we characterise, levitate, and then feedback cool. The diamagnetic particles are mesocarbon nanobeads.
The novelty of our design lies in our addition of an insulating silica coating to each particle, and the use of a high vacuum-compatible insulating wax to bind each now-insulating particle together. Together, these two improvements drastically reduce the effects of eddy current damping, and hence increase the Q-factor.  

\begin{figure}[ht] 
		\includegraphics[width=8cm]{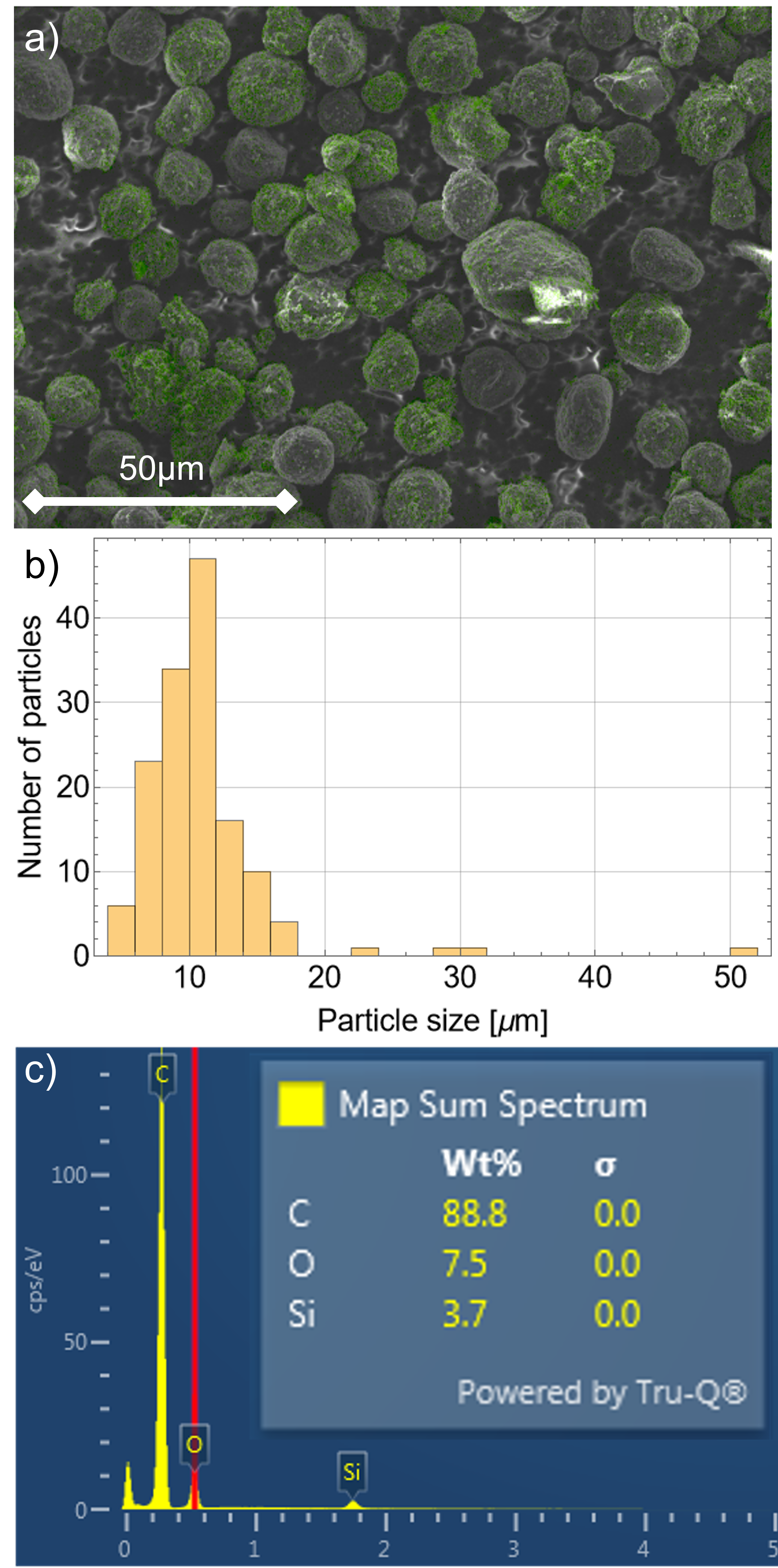} \\
		\centering
		\caption{a) A Scanning Electron Microscopy (SEM) image of the coated graphite. Overlaid is an Energy-Dispersive X-ray (EDX) mapping showing the presence of silicon in green. b) Histogram of graphite particle sizes. These are predominantly distributed around 10 micrometers. c) The EDX spectrum showing relative proportions of carbon, oxygen and silicon. The presence of silicon can only come from the applied silica coating.}
		\label{Silica_EDX}
\end{figure}

The graphite particles used are Nanografi Micron Powder which is primarily used for Lithium Ion Battery construction (NG08BE0307). An example SEM image, size histogram, and elemental analysis is shown in \cref{Silica_EDX}. The average diameter of the particles is $11 \pm2\:\mathrm{\mu m}$.

We coat the graphite particles with silica in a sol-gel process, as detailed in the next section. Once coated the powder is mixed with 
Apiezon\textsuperscript{\tiny\textregistered} Wax W, a vacuum-compatible wax in 3:2 ratio by mass. The wax is crushed into small pieces, which are thoroughly mixed with the coated graphite powder. The mixture is melted on an AS ONE CHPS-170DN hotplate at a temperature of $150^\circ\mathrm{C}$. Next, it is rolled out on a silicone mat with thickness guides on either side of the soft wax. A stainless steel rod is used as a rolling pin to produce a uniformly thick resonator. After the composite material has cooled down, the resonator is cut into a square. The resonator's dimensions are $(7.88 \pm 0.05) \rm mm \times (7.97 \pm 0.01)\rm mm \times (0.530 \pm 0.008) mm$. By volume, the composite material is 43.3\% graphite powder while by mass it is 60\%.

\subsection{Particle Coating}
\begin{figure}[ht] 
		\includegraphics[width=8cm, height= 2.4cm]{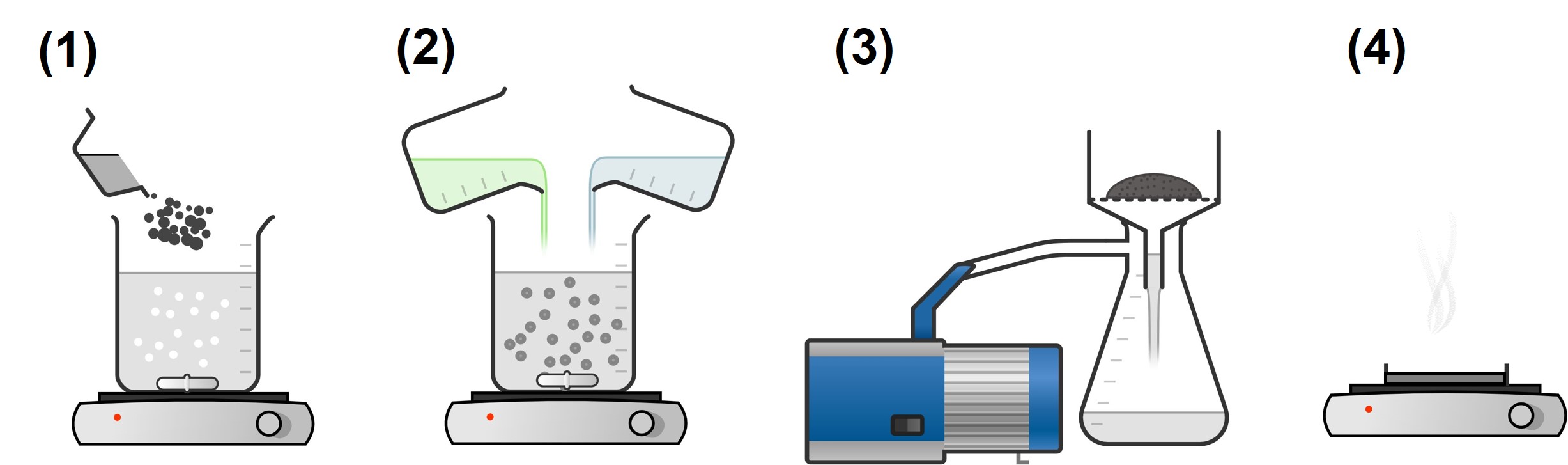}
		\centering
		\caption{Schematic diagram for the silica coating of the graphite particles. First, the graphite is poured into ethanol, to which the PEG is added and agitated at 500 rpm for 6 hours. After this, the TEOS is added with ammonium hydroxide, which is agitated for 17 hours. Then the solution is filtered and washed with ethanol, and left to dry at 60 $^\circ$C on a hot plate for 2 hours.}
		\label{coating_procedure}
\end{figure}

The silica coating on the graphite was prepared by a one-step process through physical adsorption of PEG\cite{Kim2017AGraphite} and is illustrated in \cref{coating_procedure}.
\begin{enumerate}
    \item A quantity of 5 grams of graphite was poured into a solution of 75 mL of ethanol (Sigma-Aldrich, 99.5\%) containing 0.5 grams of PEG 4600 (Aldrich). The resulting mixture was agitated at a speed of 500 rpm for a duration of 6 hours at room temperature.
    \item Once the PEG had been adsorbed onto the graphite, a solution of ammonium hydroxide (Sigma-Aldrich, 25\% solution in water), which is used as a catalyst for the sol-gel reaction, and TEOS (Aldrich, 99.0\%), which served as the silica precursor, was added to the mixture in a ratio of 1:3. This sol-gel reaction was stirred at room temperature for a period of 17 hours on the hot plate.
    \item Following the coating process, the mixture was filtered and washed with ethanol.
    \item Subsequently, the moist filter cake was dried on the hot plate at a temperature of 60 $^\circ$C for 2 hours. 
\end{enumerate}   
\begin{figure}[ht] 
		\includegraphics[width=5cm, height= 2.77cm]{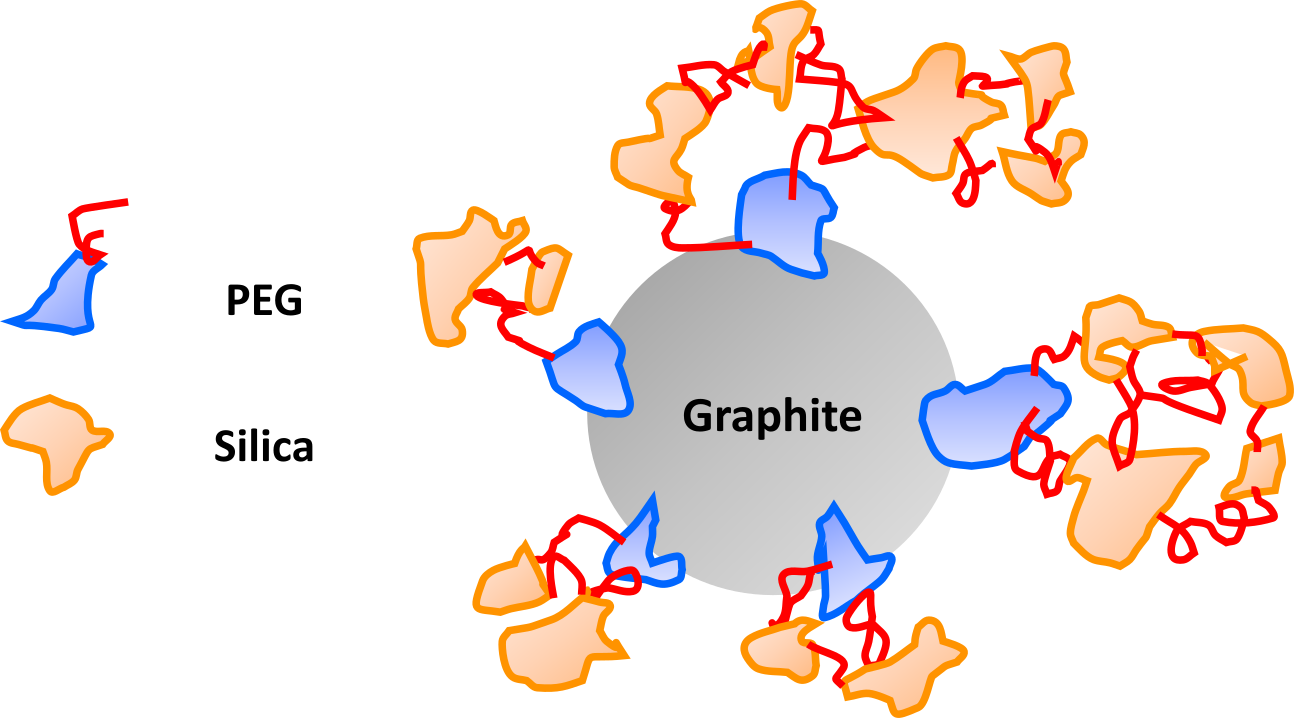}
		\centering
		\caption{Diagram illustrating the mechanism by which the graphite powder is coated with silica. The PEG is first adsorbed onto the graphite. Subsequently, silica is generated and attached to the PEG via the silica precursor TEOS and ammonium hydroxide.}
		\label{coating_mechanism}

\end{figure}

The silica coating was confirmed by using Scanning Electron Microscope (SEM) imaging and elemental analysis via Energy Dispersive X-ray (EDX) spectroscopy. The mapping of silicon from the EDX analysis is shown in Fig. \ref{Silica_EDX}. The inset in Fig. \ref{Silica_EDX}(c) shows a typical map sum spectra, showing that by weight, 3.7 \% of the sample imaged is the element silicon- which can only be due to the silica coating. 

\subsection{Electrical Conductivity}

When a conductive material such as HOPG moves through a magnetic field, eddy currents are produced in the material which causes strong damping. The resistance of our novel composite resonator was measured with the Hioki RM3545 Resistance meter. The Hioki RM3545 can measure resistances in the range of 10 m$\Omega$  to 1 G$\Omega$ with an accuracy of 0.006\% rdg. However, the resistance meter was still unable to give a resistance measurement, indicating that the resistance of the resonator is above the measurable range. This corresponds to a maximum possible conductivity of 8.6 x $10^{-9}$ S/m. The Apiezon Wax itself has a conductivity of 1.59 x $10^{-16}$ S/m. In comparison, the conductivity of a resonator made with uncoated graphite powder was found to be on the order of 10$^{-1}$ S/m. This shows that the silica coating on the graphite substantially suppresses current flow.

\subsection{Measurement of Diamagnetic Susceptibility}

The diamagnetic susceptibility of HOPG is extremely anisotropic, with susceptibility $\chi_\perp\sim 450 \times 10^{-6}$ perpendicular to the graphite planes, and $\chi_\|\sim 85 \times 10^{-6}$ parallel to them. For solid HOPG resonators this anisotropy is exploited, and the flat resonator is cut such that the larger susceptibility is normal to the plane of the resonator. Using a composite of wax, silica and graphite beads instead of a HOPG slab naturally results in a decreased diamagnetic susceptibility of the resonator, due to the random orientation of the beads. This decrease in susceptibility, combined with the lower fraction of graphite in the resonator, leads to a decrease in levitation height. 
We note that the literature reports that the diamagnetic susceptibility of graphite samples drops off rapidly for thicknesses below 30 $\mu$m. \cite{Semenenko2018DiamagnetismRevised}.

We conducted measurements of the magnetic susceptibility of the graphite powders using a Vibrating Sample Magnetometer (VSM) in the  PPMS (P525) VersaLab (V525) system. The sample is placed in a constant magnetic field. The magnetic dipole moment of the sample in that field creates a magnetic field. As the sample is moved up and down the resultant magnetic field changes as a function of time, inducing an electric field in the VSM's pickup coils. The induced current is proportional to the magnetization of the sample. 

We measured a 0.006 g sample of the Nanografi graphite powder in the VSM, and the magnetic moment was measured for fields in the range of $\pm$5 T. The susceptibility of the graphite powder is found to be $\chi$ = -218 $\times$ 10$^{-6}$. From this, we estimate the average susceptibility of the resonator plate to be - 90.2 $\times$ 10$^{-6}$ \textcolor{black}{from the volume fraction in the composite}.

\section{Modelling of Damping of Insulating Resonators}\label{Section:modelling_damping}

To understand the expected eddy damping of the composite resonator, a finite element methods (FEM) model was built in COMSOL. 
In the model, we assume that in the composite resonator one has spherical coated graphite particles which are evenly distributed in the insulating Apiezon wax matrix. We consider the graphite spheres to have a uniform diameter of $12 \:\mu$m. Pyrolytic graphite has an anisotropic susceptibility with different in-plane and out-of-plane components ($\chi_{||}=-85\times 10^{-6}$ and $\chi_{\perp}=-450\times 10^{-6}$ ). However, in the composite, the graphite spheres are randomly oriented in the matrix which leads to an effective isotropic susceptibility $\chi_{eff}$ of the composite resonator. We note that the susceptibility of graphite is size dependent, and it will decrease with reducing particle size\cite{Semenenko2018DiamagnetismRevised}. %. 
For the $12\: \mu m$ size particle, the vertical component of susceptibility can drop to half of that of bulk graphite. If we adopt this assumption we can roughly estimate the isotropic effective susceptibility as $\chi_{eff}=\frac{1}{3}\times(\frac{1}{2}\times\chi_{\perp})+\frac{2}{3}\times\chi_{||}=-132\times 10^{-6}$, but this does not agree with our experimentally measured value of $\chi_{eff}\sim -218 \times 10^{-6}$. A straightforward average of $\chi_{eff}=\frac{1}{3}\times\chi_{\perp}+\frac{2}{3}\times\chi_{||}=-207 \times 10^{-6}$, agrees with our measurements, and we use this for our damping FEM simulations. 
The effective conductivity of the randomly oriented graphite sphere can be estimated as $\sigma_{eff}=\frac{1}{3}\times\sigma_{\perp}+\frac{2}{3}\times\sigma_{||}=1.33\times 10^{5} \rm S/m$.
Due to the symmetries of the setup, we can model 1/8 of the resonator for vertical motion to reduce the computation time and memory needed. We select around 500 evenly distributed spheres in one layer to represent all the graphite particles in the composite. By simulating the Lorentz force on the resonator when it oscillates around the equilibrium position, we calculate the damping coefficient and {\color{black} get Q factors of the motion $Q\sim 5\times 10^{6}$}. The Q from the simulation is almost one order of magnitude higher than the measured Q for our resonator, which could be attributed to that there is a distribution of particle sizes in the real resonator and the eddy damping is dominated by a few large particles.

\begin{figure}[h!]
\begin{center}	
\includegraphics[width=\linewidth]{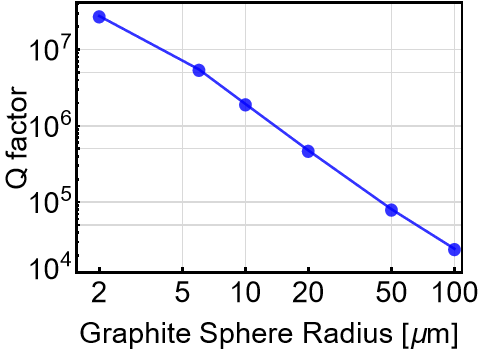}
\end{center}
\vspace*{-10pt}
\caption{FEM(Comsol) simulation of the Q factor due to eddy current damping as a function of the radius of the graphite spheres. The total masses of the graphite spheres are kept fixed for the simulation of spheres with different sizes.}
    \label{Q_simulation}
\end{figure}

\section{Measurement of Quality Factor}\label{Section:qualfactor}

\begin{figure}[h!]
\begin{center}	
\includegraphics[width=\linewidth]{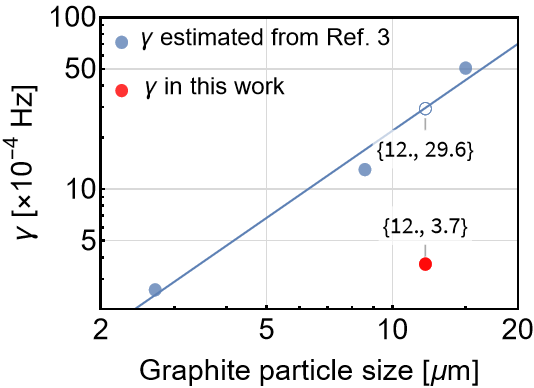}
\end{center}
\vspace*{-10pt}
\caption{Comparison of damping rate $\gamma$ in this work with \citeinline{Chen2020} at the pressure of $\sim 10^{-6} \rm mBar$. 
The red circle shows the damping $\gamma_{12\,\mu{\rm m}} \sim 3.7\times 10^{-4} \:\rm Hz$ of our the resonator. The median particle size was measured to be $12 \: \mu \rm m$ via Scanning Electron Microscopy. The blue circles \textcolor{black}{are estimated from} Ref. \cite{Chen2020}. The empty circle denotes a linear fitting with  $\gamma_{12\,\mu{\rm m }} \sim 29.6\times 10^{-4} \:\rm Hz$.
The insulating coating effectively suppresses the damping by almost one order of magnitude.
}
    \label{QComparison}
\end{figure}

We measured the quality factor of our composite plate using ringdown measurements. 
This experimental approach entailed a controlled electromagnetic excitation of the composite plate at its resonance frequency, followed by the cessation of the excitation. 
Subsequently, we recorded the ringdown decay and captured the system's response as it settled into its damped state.
The decay of the oscillations exhibited exponential behavior which we fit as  $x(t) =A e^{-\gamma t}$, where the decay constant is $\gamma = \pi f_{0}/Q$. 
We performed ringdown measurements at high vacuum ($P=1.2\times 10^{-6} {\rm mBar}$). The velocity was fit to exponential decay, yielding $\gamma=3.7\times 10^{-4} \rm Hz$ and $Q=1.58\times 10^{5}$. As shown in \cref{QComparison}, the damping is almost one order of magnitude lower than that of the composite in \citeinline{Chen2020} with the same particle size. This shows that the insulating coating on the graphite particle can effectively block the conduction between particles and suppress the overall eddy damping.   

\section{Orientation of Resonators:}\label{Section:orientation}

In this section, we detail the theoretical study of the equilibrium configuration of the graphite and composite plate above the checkerboard magnet array. We will find that for HOPG the equilibrium orientation is angled at $\pi/4$ relative to the magnets. The composite sits aligned with the checkerboard array. This agrees with the experimental observations. 

From symmetry, we will assume that the plate's centre of mass is positioned along the $z$-axis. We now derive and study the potential energy of the plate, taking into account both the gravitational and magnetic energies as a function of the centre of mass height and azimuthal angle $\phi$. The equilibrium configurations will occur at the potential minima. We first study the magnetic fields generated by the checkerboard magnet array.

\subsection{Magnetic field of checkerboard array}\label{app:CheckerboardField}
The magnetic field of a cuboidal magnet is given in \S 2.5 of \citeauthor{Camacho2013}.\cite{Camacho2013} Suppose the cube has side lengths $D$, and is centered on the origin with magnetization oriented along the $+\hat{z}$ direction. We first define two helper functions:
\begin{equation}
    \begin{aligned}
        F_1 &= \mathrm{arctan}\frac{(\tilde{x}+1/2)(\tilde{y}+1/2)}{(\tilde{z}+1/2)\sqrt{(\tilde{x}+1/2)^2+(\tilde{y}+1/2)^2+(\tilde{z}+1/2)^2}}, \\
        F_2 &= \frac{\sqrt{(\tilde{x}+1/2)^2+(\tilde{y}-1/2)^2+(\tilde{z}+1/2)^2}+1/2-\tilde{y}}{\sqrt{(\tilde{x}+1/2)^2+(\tilde{y}+1/2)^2+(\tilde{z}+1/2)^2}-1/2-\tilde{y}},
    \end{aligned}
\end{equation}
where $(\tilde{x},\tilde{y},\tilde{z})=(x,y,z)/D$. 
The magnetic field from the cube can then be written as:
\begin{equation}\label{eq:NDMagneticField}
    \mathbf{B}(\mathbf{r})=\mu_0M\,\tilde{\mathbf{B}}(\mathbf{r}/D),
\end{equation}
 where $\tilde{\mathbf{B}}$ is the magnetic field for a cube with unit side lengths,   
nondimensionalised by $\mu_0M$:
\begin{equation}\label{eq:BComponentsNonDim}
    \begin{aligned}
        \tilde{B}_x(\tilde{x},\tilde{y},\tilde{z}) &= \frac{1}{4\pi}\ln\frac{F_2(-\tilde{x},\tilde{y},-\tilde{z})F_2(\tilde{x},\tilde{y},\tilde{z})}{F_2(\tilde{x},\tilde{y},-\tilde{z})F_2(-\tilde{x},\tilde{y},\tilde{z})}, \\
        \tilde{B}_y(\tilde{x},\tilde{y},\tilde{z}) &= \frac{1}{4\pi}\ln\frac{F_2(-\tilde{y},\tilde{x},-\tilde{z})F_2(\tilde{y},\tilde{x},\tilde{z})}{F_2(\tilde{y},\tilde{x},-\tilde{z})F_2(-\tilde{y},\tilde{x},\tilde{z})}, \\
        \tilde{B}_z(\tilde{x},\tilde{y},\tilde{z}) &= -\frac{1}{4\pi}\left[F_1(-\tilde{x},\tilde{y},\tilde{z})+F_1(-\tilde{x},\tilde{y},-\tilde{z})\right. \\
            &\phantom{=\frac{1}{4\pi}[[+} +F_1(-\tilde{x},-\tilde{y},\tilde{z})+F_1(-\tilde{x},-\tilde{y},-\tilde{z}) \\
            &\phantom{=\frac{1}{4\pi}[[+} +F_1(\tilde{x},\tilde{y},\tilde{z})+F_1(\tilde{x},\tilde{y},-\tilde{z}) \\
            &\phantom{=\frac{1}{4\pi}[[+} \left.+F_1(\tilde{x},-\tilde{y},\tilde{z})+F_1(\tilde{x},-\tilde{y},-\tilde{z})\right].
    \end{aligned}
\end{equation}
The magnetic field of a magnet centered at $\mathbf{r}_0$ is given by $\mathbf{B}(\mathbf{r}-\mathbf{r}_0)$. 

The checkerboard array is constructed by placing four magnets centered at $(-D/2,-D/2)$, $(-D/2,D/2)$, $(D/2,D/2)$, and $(D/2,-D/2)$, with magnetizations alternating between $\pm M$. The sum of the corresponding magnetic fields gives us the field of the array.

\subsection{Potential energy of the plate}
We will now find an expression for the potential energy of the plates. Graphing this as a function of levitation height and rotation angle will allow us to study their orientation. We consider the plate to have a width of $L$, and thickness $\delta$.
We take the $z$-component of the diamagnetic susceptibility to be $\chi_{z}$, and suppose that the $x$- and $y$-components take the same value of $\chi_{xy}$. Off-diagonal elements of the susceptibility tensor are taken to be zero. The plate lies parallel to the $x,y$-plane, with centre of mass along the $z$-axis at $\vec{r}=(0,0,z)$. Thus $z$ denotes the position of the centre of the plate, while the upper surface will be located at $z+\delta/2$. We note that in \citeinline{Chen2022DiamagneticResonators} $z$ was used to refer to the top of the plate, which must be taken into account if one wishes to compare our values with theirs. By $\phi$ we will denote the rotation of the plate about the $z$-axis. We will take the orientation at $\phi=0$ to align with the magnets. Due to rotational symmetry, this will also occur at $\phi=n\pi/2$ for integer $n$.

We will first consider the magnetic contribution to the potential energy. This is the integral of the diamagnetic energy density $-\chi_j|B_j|^2/2\mu_0$ over the plate's volume, where $B_j$ are the components of $\mathbf{B}$ defined in \cref{eq:NDMagneticField}. We will denote this volume $\mathcal{V}(L,z,\phi)$, which depends on the length, height, and rotation of the plate. The magnetic energy is then
\begin{equation}\label{eq:MagneticPotentialEnergy}
    \begin{aligned}
    U_B(L,z,\phi) &= -\frac{\delta}{2\mu_0}\iint_{\mathcal{V}(L,z,\phi)}\left[\chi_{xy}\left(|B_x|^2+|B_y|^2\right)\right. \\
    &\hphantom{=-\frac{1}{2\mu_0}}\left.+\chi_z|B_z|^2\right]\,\mathrm{d}x\mathrm{d}y,
    \end{aligned}
\end{equation}
where $\delta$ is the thickness of the plate, which is sufficiently thin that the magnetic field is approximately the same at the top and bottom of the plate. This approximation makes analysis easier, and we have verified via numerical simulation that this holds true for the thicknesses of our resonators.

The gravitational potential energy of the plate is independent of orientation $\phi$ as the plate sits in the $x-y$ plane. If $\rho$ is the mass density per unit volume and $g$ gravity, this is
\begin{equation}\label{eq:GravitationalPotentialEnergy}
    U_g(L,z)=(\rho L^2\delta)gz.
\end{equation}
Summing these gives the total potential energy:
\begin{equation}\label{eq:TotalPotentialEnergy}
    U(L,z,\phi)=U_B(L,z,\phi)+U_g(L,z).
\end{equation}
The potential energy depends on the parameters $L,\delta, z,\phi$, which describe the position, shape, and orientation of the plate. It also depends on the susceptibility $\chi_z,\chi_{xy}$ of the plate, and the physical parameters $D,M$ of the magnet.

We can reduce the number of physical parameters in \cref{eq:TotalPotentialEnergy} by expressing it in natural units, which we will denote with tildes. First, we parameterise distances as relative to the magnet length: 
$\tilde{L}=L/D$, $\tilde{z}=z/D$, $\tilde{\delta}=\delta/D$.
We next non-dimensionalize the magnetic susceptibility. For the HOPG, $\chi_z$ is significantly larger than $\chi_{xy}$. In the composite however all three are equal due to the random orientation of the particles. Thus let us define $\chi_{z0}$ to be the value of $\chi_z$ for HOPG, and in terms of this we define the dimensionless parameters:
\begin{equation}
        \tilde{c} = \chi_z/\chi_{z0},\; \tilde{\chi}_{xy} = \chi_{xy}/\chi_z.
        %=\frac{\chi_{xy}}{\tilde{c}\chi_{z0}}.
\end{equation}
Both of these are positive. The relative strength of the magnetic susceptibility is described by $\tilde{c}$, and the orientation by $\tilde{\chi}_{xy}$. For HOPG the susceptibility has maximum strength $\tilde{c}=1$, and it is strongly oriented: $\tilde{\chi}_{xy}\approx 0.19$. Due to both the random orientation of the particles and the lower density of graphite the composite has a lower susceptibility $\tilde{c}\approx 0.27$, and no orientation: $\tilde{\chi}_{xy}=1$.

Let us now express the magnetic potential energy \cref{eq:MagneticPotentialEnergy} in natural units. The integrand is
\begin{equation}
    \begin{aligned}
        \phantom{=}&\chi_{xy}\left(|B_x|^2+|B_y|^2\right)+\chi_z|B_z|^2 \\
            &=(\mu_0M)^2\left[\chi_{xy}\left(|\tilde{B}_x|^2+|\tilde{B}_y|^2\right)+\chi_z|\tilde{B}_z|^2\right], \\
            &=(\mu_0M)^2\chi_z\left[\tilde{\chi}_{xy}\left(|\tilde{B}_x|^2+|\tilde{B}_y|^2\right)+|\tilde{B}_z|^2\right],
    \end{aligned}
\end{equation}
where the dimensionless magnetic field components $\tilde{B}_j$ were defined in \cref{eq:BComponentsNonDim}. Motivated by this we define the dimensionless magnetic potential energy as
\begin{equation}
    \begin{aligned}
    \tilde{U}_B(\tilde{L},\tilde{z},\phi) &= \frac{1}{2}\iint_{\tilde{\mathcal{V}}(\tilde{L},\tilde{z},\phi)} \left[\vphantom{\int}\tilde{\chi}_{xy}\left(\lvert\tilde{B}_x\rvert^2+\lvert\tilde{B}_y\rvert^2\right) +\lvert\tilde{B}_z\rvert^2\vphantom{\int}\right]\,\mathrm{d}\tilde{x}\,\mathrm{d}\tilde{y}.
    \end{aligned}
\end{equation}
Note that in contrast to \cref{eq:MagneticPotentialEnergy}, the sign in the dimensionless energy is positive. This is to account for the fact that $\tilde{\chi}_{xy}$ is positive, whereas $\chi_{xy}$ and $\chi_z$ are negative. Using the chain rule $\mathrm{d}x\,\mathrm{d}y=D^2\mathrm{d}\tilde{x}\,\mathrm{d}\tilde{y}$, we derive the relation:
\begin{equation}
    U_B = \frac{\delta}{\mu_0}\left[(\mu_0M)^2\lvert\chi_z\rvert\right]D^2\tilde{U}_B=\mathcal{E}\,\tilde{c}\tilde{U}_B,
\end{equation}
where we've defined our natural energy scale:
\begin{equation}
    \mathcal{E}=\lvert\chi_{z0}\rvert\mu_0M^2\delta D^2.
\end{equation}

Let us pause to investigate the meaning of the energy scale $\mathcal{E}$. This is a positive number, due to the absolute value around $\chi_{z0}$. The energy density of the HOPG is given by $\chi_{z0}|B|^2/(2\mu_0)=|\tilde{B}|^2\chi_{z0}\mu_0M^2$. The quantity $\delta D^2$ corresponds to the volume of a graphite plate with side length $D$. The natural energy scale can thus be thought of as the diamagnetic potential energy of a graphite slab with side lengths equal to those of the magnets.

We can also write the gravitational potential energy in terms of $\mathcal{E}$:
\begin{equation}
    U_g=\left(\delta D^2\mu_0M^2\lvert\chi_{z0}\rvert\right) \tilde{L}^2\left(\frac{\rho gD}{\mu_0M^2\lvert\chi_{z0}\rvert}\right)\tilde{z}.
\end{equation}
Defining the effective gravity
\begin{equation}
    \tilde{g}=\frac{\rho gD}{\mu_0M^2\lvert\chi_{z0}\rvert},
\end{equation}
the gravitational potential energy is then
\begin{equation}
    U_g=\mathcal{E}\tilde{L}^2\tilde{g}\tilde{z}.
\end{equation}
Putting these together, the dimensionless potential energy $\tilde{U}=U/\mathcal{E}$ is given by
\begin{equation}\label{eq:tildeU}
    \tilde{U}(\tilde{L},\tilde{z},\phi)=\tilde{c}\tilde{U}_B(\tilde{L},\tilde{z},\phi)+\tilde{g}\tilde{L}^2\tilde{z}.
\end{equation}
An immediate consequence of this analysis is that the energy, and hence position and orientation, are independent of the thickness of the plate. 

\begin{figure*}
\begin{center}	
\includegraphics[width=\linewidth]{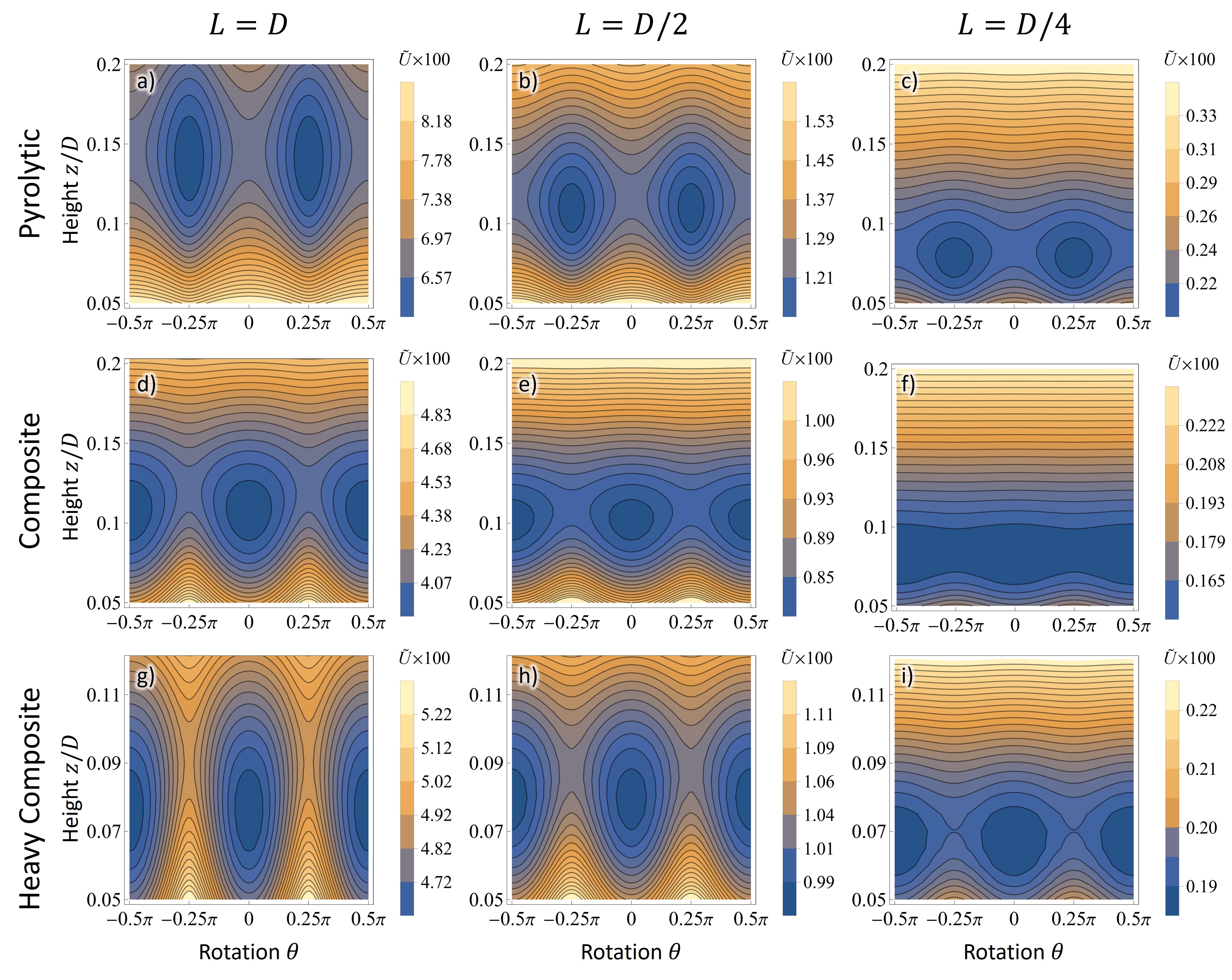}
\end{center}\vspace*{-10pt}\caption{Difference in orientation and levitation height between the pyrolytic and composite graphite plates. We plot the dimensionless potential energy $\tilde{U}$ from \cref{eq:tildeU}, with axes corresponding to the rotation angle $\theta$, and the centre-of-mass height $z$ relative to the magnet lengths $D$. Columns denote the length $L$ of the slab, relative to the magnet length. The upper row shows the pyrolytic graphite, and the middle row is the composite. The pyrolytic graphite has a density $\rho=2700\mathrm{kg}/\mathrm{m}^3$, magnetic susceptibility $\chi=-(85,85,450)\times10^{-6}$. The composite has density $\rho=1442\mathrm{kg}/\mathrm{m}^3$, and susceptibility $\chi=-(120,120,120)\times10^{-6}$. To compare solely the magnetic effects, the lower row shows a composite with density equal to that of pyrolytic graphite. This levitates at a lower height, thus the vertical axis of this plot is half that of the other plots. For pyrolytic graphite the equilibrium orientation is always at $\pm \pi/4$, while for the composite it is $\pm \pi/2,0$. As slab length decreases the equilibrium height tends to decrease. We note that the trapping regions for the composite are much shallower than those of the pyrolytic graphite, as indicated by the size of the contours at minimum energy.}
\label{fig:3d-potential}
\end{figure*}

We plot the energy in \cref{fig:3d-potential}. From the experimental and the numerical analysis it is clear that the anisotropy of the effective averaged diamagnetic susceptibility plays a strong role in the equilibrium configuration of the levitated slab. For HOPG the orientation of the slab is angled at $\phi_{\rm HOPG}=\pm \pi/4$, while for the composite it is angled at $\phi_{\rm Comp}=\,\pm \pi/2$. As discussed in the main text these results appear to disagree with observations made in \citeinline{Chen2022DiamagneticResonators}. We speculate that this is likely due to their magnets having geometry different from perfect cubes, or possibly some alignment of the graphite particles in the resin as it cured.
 
\twocolumngrid\
%%%%%%%%%%%%%%%%%%%%%%%%%%%%%%%%%%%%

\section{Description of Experimental Setup}\label{Section:setup}

\begin{figure}[h] 
		\includegraphics[width=8cm, height= 3.8cm]{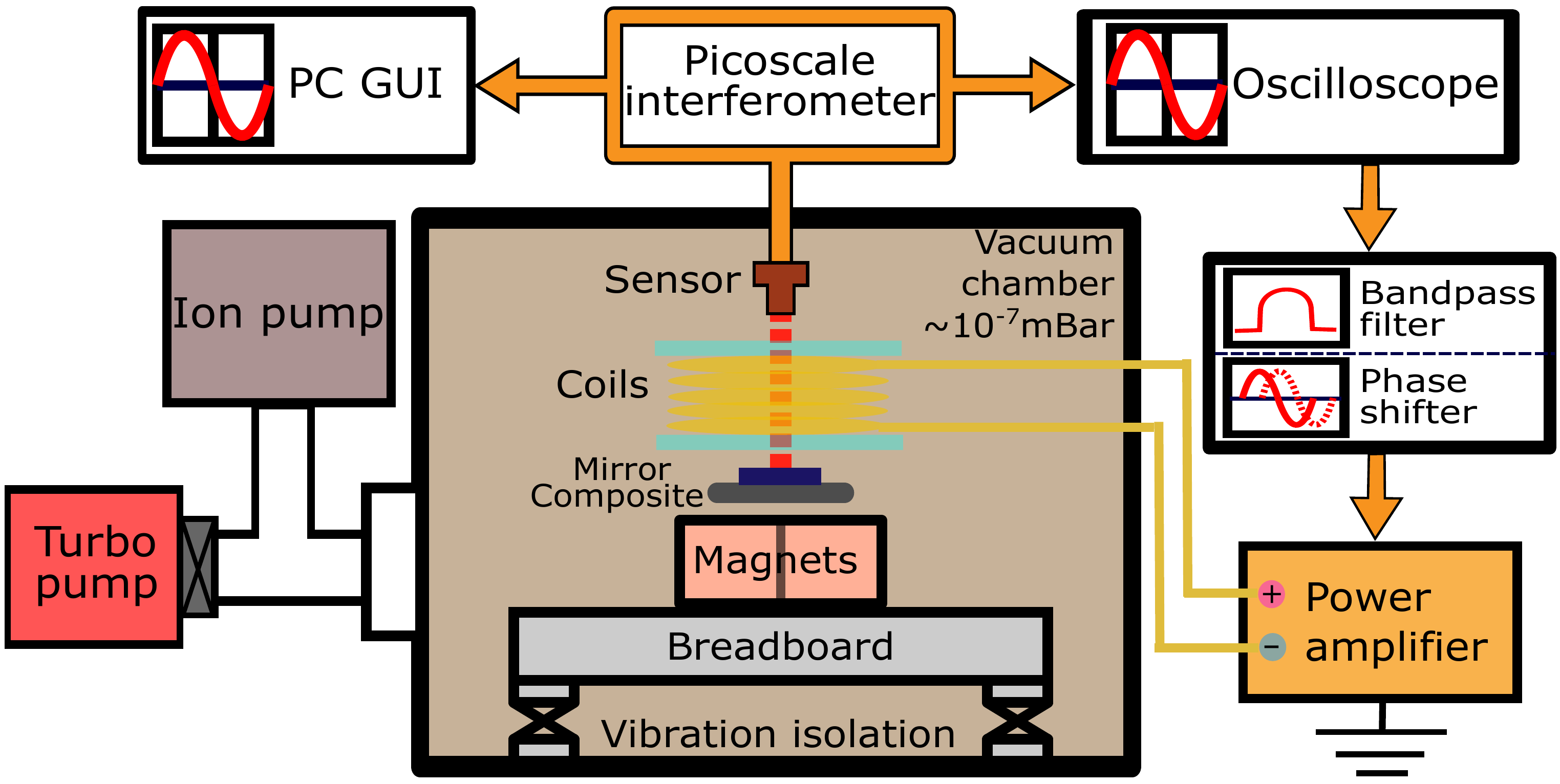}
		\centering
		\caption{Schematic of experimental setup. The novel insulating resonator is levitated by four NdFeB N52 magnets, which are fixed on a breadboard resting upon four vibration isolation mounts within the vacuum chamber. The vacuum chamber and ion pump lie on a vibration isolation optical table, while the turbopump is on a separate vibration isolation platform, and is only used for initial pump-down, and switched off during cooling and Q-factor measurements, while the vibrationless ion pump (on the same optical table as the experiment) maintains high vacuum. A small mirror is fixed to the graphite sample for reflecting laser to interferometrically measure the position. The output position (or velocity) signal is sent to a digital bandpass filter which only allows through the component of the vertical mode. The resulting signal with a specific phase shift is then fed to the coil through a power amplifier, which applies magnetic force to the resonator. }
		\label{Setup_diagram}
\end{figure}

The setup used for the results presented here is shown in Fig. 1 (Main manuscript), which we reproduce here in  \cref{Setup_diagram}. The insulating resonator is levitated by an alternating polarity checkerboard magnet array made up of four NdFeB magnets. They are rigidly held within a holder that is mounted on a small optical breadboard. The breadboard itself sits on four vibration isolation supports. The velocity and displacement of the resonator are monitored by an interferometric displacement sensor (SmarAct PICOSCALE Interferometer). It is based on a compact Michelson interferometer, and enables high-precision measurements in real time with a resolution of picometers at a high bandwidth. The sensor of the PICOSCALE is fixed to a five-axis ultra-high vacuum-compatible motorised stage. The whole structure is placed in a vacuum chamber, which is evacuated by a system consisting of a turbopump, an associated roughing pump, and an ion pump. During the measurement periods, the turbopump is switched off to avoid mechanical vibrations, while the ion pump operates continuously to maintain a high vacuum. To isolate the setup from vibrations, the vacuum chamber and ion pump are supported by a damped and vibration-isolated optical table, and the turbopump is supported by a separate vibration-damped and isolated platform.

In addition to the above equipment, a Field Programmable Gate Array (FPGA) is used. 
An FPGA is a semiconductor device that can be reprogrammed to perform complex computational tasks, offering a balance between the general-purpose flexibility of a processor and the high performance and efficiency of an Application-Specific Integrated Circuit (ASIC). The FPGA used for signal processing is the Red Pitaya (STEMlab 125-14). It is mainly used for the bandpass filter to isolate the vertical mode signal from noise. In addition, it is used to create some delay in the feedback signal for reducing the phase difference with the input signal for optimal cooling and studying the effect of the delayed feedback cooling.

The bandpass filter that we have used is a conventional single-rate finite impulse response (FIR) filter which has a characteristic linear phase response. The length of the filter coefficients used is 1001, calculated using the Hamming window method. To attenuate peaks around $f\sim$ 16.2 Hz and 25 Hz, we have decimated the sampling frequency to 1250 Hz and set low and high passband frequencies to 18 Hz and 23 Hz. The amplitude and phase response of the bandpass filter can be seen in \cref{fig:Filter_Signal_Response}.

\section{Characterization of Feedback Force}\label{Section:Feedback_force}

We shift the equilibrium height of the resonator by applying an additional magnetic field through a multiturn solenoid coil mounted directly above the resonator. This actuation is how we physically implement feedback control of the resonator.

Our actuation solenoid has an inner radius of $5\:\rm mm$ and height of $10\:\rm mm$, and consists of $368$ turns of AWG24 copper wire. To calibrate the resonator's response to actuation, we simulate the magnetic force on and displacement response of the resonator from a DC current through the coil, using the FEM package COMSOL. 

\begin{figure}[h!]
\begin{center}	\includegraphics[width=\linewidth]{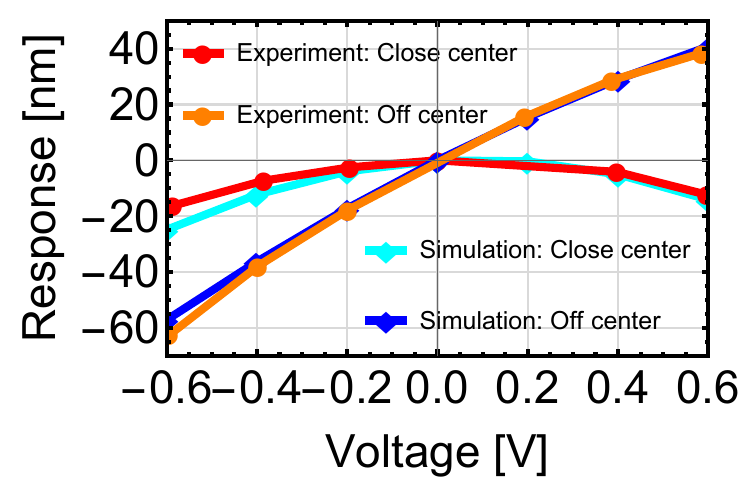}
\end{center}
\vspace*{-10pt}
\caption{The vertical displacement response of the composite resonator to the driving voltage of the actuation coils. The vertical displacement \textcolor{black}{response} of the resonator is symmetric with the applied voltage if the coil axis is vertically and co-axially aligned well with the center of the resonator. However,  the displacement response is nearly linear with the applied voltage if the coil axis is non-coaxial with the resonator center, where it is displaced horizontally along the diagonal of the magnet.}
    \label{Coil_Force}
\end{figure}

The simulation results show good agreement with experimental observation (the blue and red/orange curves in Figure \ref{Coil_Force}, respectively). When the axis of the solenoid is aligned to the center of the resonator and the magnet array, the displacement of the resonator is symmetric with respect to the sign of the current. In this case, the force density along the $k-$axis, $f_k\sim \nabla |B_k|^2$, 
is the same regardless of the direction of the magnetic field generated by the coil  (direction of the current in the coils). The displacement-current response 
change to be more monotonic-like if the axis of the coils is shifted diagonally away from the center of the resonator but for the cooling we arrange the coil and resonator to be co-axial.
During the feedback cooling experiment, we impart a DC shift on the feedback signal to make sure the feedback operates away from $V=0 {\rm V}$, i.e. within the monotonic response regime.

\section{Theory and Simulations of Delayed Feed-back Cooling}\label{Section:feedback_theory_simulations}

To understand the measurement signal and design feedback schemes, in this section we will derive and analyze a simple model of the system. The interferometric measurement and coil-based actuation in our experiment will most effectively interface with the vertical motion of the plate. We will approximate this vertical motion as harmonic, and uncoupled from the other rigid-body translational and rotation modes. This approximation should be valid for small excitations, and when the plate is driven close to the resonant frequency of this mode. 

\subsection{Equations of motion}
We will approximate the system as a one-dimensional harmonic oscillator with thermal driving. This has an equation of motion
\begin{equation}\label{eq:DampedOscillator}
    \ddot{x}(t)+\gamma\,\dot{x}(t)+(2\pi f_0)^2x(t)=\sqrt{\frac{2\gamma k_BT}{m}}\xi(t),
\end{equation}
where a dot represents the time derivative. The oscillator has natural frequency $f_0$, damping frequency $\gamma$, and mass $m$. The temperature of the thermal bath is $T$, and $k_B$ is Boltzmann's constant. The bath is modeled as Gaussian white noise, denoted $\xi(t)$. This is a randomly fluctuating process. Deterministic values can be extracted only by taking expectation values, and using the mean and correlation function:
\begin{align}
    \langle\xi(t)\rangle &= 0, \\
    \left\langle\xi(t_1)\xi(t_2)\right\rangle &= \delta(t_1-t_2).\label{eq:DeltaCorrelation}
\end{align}
Since the Kronecker function $\delta(t)$ has units of inverse time $1/t$, the correlation function \cref{eq:DeltaCorrelation} implies $\xi(t)$ has units $1/\sqrt{t}$. The intuition behind this is that $\xi(t)\Delta t$ represents the distance traveled by a particle undergoing Brownian motion in a short time interval $\Delta t$. For ideal Brownian motion, this is proportional to $\sqrt{\Delta t}$, as shown by Einstein.

We can reduce the number of physical parameters by moving to natural units for this system. Suppose we have a natural length scale $\ell$, and frequency $\nu$. Using these we can define a dimensionless time and position
\begin{equation}
    \tilde{t}=\nu t,\;\tilde{x}=x/\ell,
\end{equation}
where dimensionless quantities are denoted with a tilde. This change of timescale manifests as a scaling factor in the white noise. We'll denote $\xi(\tilde{t})$ as $\tilde{\xi}$. Then since $\tilde{\xi}$ is dimensionless (since $1/\sqrt{\tilde{t}}$ is dimensionless), we must have
\begin{equation}
    \tilde{\xi}(\tilde{t})=\frac{1}{\sqrt{\nu}}\xi(t).
\end{equation}
Note that moving to natural units does not change the correlation function. To see this, note that the Kronecker function scales as $\delta(ax)=\delta(x)/a$, which implies
\begin{equation}\label{eq:NDNoiseCorrelation}
    \begin{aligned}
        \langle\tilde{\xi}(\tilde{t}_1)\tilde{\xi}(\tilde{t}_2)\rangle &= \frac{1}{\nu}\langle\xi(t_1)\xi(t'_2)\rangle, \\ &= \frac{1}{\nu}\delta(t_1-t_2), \\ &= \delta(\tilde{t}_1-\tilde{t}_2).
    \end{aligned}
\end{equation}

Using primes to denote derivatives with respect to $\tilde{t}$, \cref{eq:DampedOscillator} expressed in natural units becomes
\begin{equation}
    \begin{gathered}
    \ell\nu^2\tilde{x}''+\gamma\ell\nu\tilde{x}'+(2\pi f_0)^2\ell\tilde{x}=\sqrt{\frac{2\gamma k_BT}{m}}\xi(t), \\
    \tilde{x}''+\frac{\gamma}{\nu}\tilde{x}'+\left(2\pi\frac{f_0}{\nu}\right)^2\tilde{x}=\frac{1}{\ell\nu^{3/2}}\sqrt{\frac{2\gamma k_BT}{m}}\frac{1}{\sqrt{\nu}}\xi(t), \\
    \tilde{x}''+\frac{\gamma}{\nu}\tilde{x}'+\left(2\pi\frac{f_0}{\nu}\right)^2\tilde{x}=\left(\frac{1}{\ell\nu}\sqrt{\frac{2k_BT}{m}}\right)\left(\sqrt{\frac{\gamma}{\nu}}\right)\left(\frac{1}{\sqrt{\nu}}\xi(t)\right). \\
    \end{gathered}
\end{equation}
This suggests that our natural length scale and frequency should be defined as
\begin{equation}
        \nu  = f_0,\;
        \ell = \sqrt{\frac{2k_BT}{m\nu^2}}.
\end{equation}
The natural timescale $1/\nu$ is equal to the oscillator period. The length-scale $\ell$ conveys the effects of the effects of temperature and mass. A higher temperature increases the amplitude of oscillations, while a larger mass or stiffer spring decreases the amplitude. We will define the dimensionless damping parameter as
\begin{equation}
    \tilde{\gamma}=\frac{\gamma}{\nu}.
\end{equation}

Expressed in natural units, \cref{eq:DampedOscillator} becomes
\begin{equation}\label{eq:DampedOscillatorND}
    \tilde{x}''+\tilde{\gamma}\,\tilde{x}'+(2\pi)^2\,\tilde{x}=\sqrt{\tilde{\gamma}}\,\tilde{\xi}.
\end{equation}
The dynamics of a harmonic oscillator with thermal driving has effectively only one parameter. This is $\tilde{\gamma}$, the ratio of the damping rate to the natural frequency. Note that the presence of $\tilde{\gamma}$ on both sides on both sides of the equations is a manifestation of the fluctuation-dissipation theorem, a relationship between the damping rate and thermal fluctuations.

\subsection{Power spectral density}
The Power Spectral Density (PSD) of a time series measures its distribution of power among different frequencies. This type of measurement is very accessible experimentally, and highly precise. In this section, we will calculate the expected PSD of the thermally driven oscillator.

The power spectral density is given by the absolute square of the Fourier transform, averaged over many realisations of the noise. This allows us to extend concepts from Fourier theory to stochastic systems, whose Fourier transform would differ with each realisation of the noise. It is necessary to take the absolute square since the phase varies randomly with the noise, and thus averaging the unsquared Fourier transform would simply yield zero. For a random variable $X(t)$, the PSD is denoted $S_{XX}$, and may be calculated as:
\begin{equation}\label{eq:PSDDefinition}
    S_{XX}(\omega)=\lim_{T\rightarrow\infty}\frac{1}{T}\int_0^T\int_0^T\mathrm{d}t_1\mathrm{d}t_2e^{-i\omega (t_1-t_2)}\left\langle X(t_1)X(t_2)\right\rangle.
\end{equation}
We can interpret \cref{eq:PSDDefinition} as the expected value of the absolute square of the Fourier transform of $X(t)$. The prefactor of $1/T$ normalises for the fact that the power in Brownian motion grows linearly with time. The units of $S_{XX}$ are thus $X^2T$. From this form we can see that the PSD of a constant multiple of $X$ is:
\begin{equation}\label{eq:PSDConstantMultiple}
    S_{(aX)(aX)}=|a|^2S_{XX}.
\end{equation}

Let us consider how the PSD is affected by our choice of natural units. Since we have chosen a natural frequency $\nu=f_0$, the dimensionless frequency $\tilde{\omega}$ will be:
\begin{equation}
    \tilde{\omega}=\omega/\nu=\omega/f_0.
\end{equation}
Then $\omega t=\tilde{\omega}\tilde{t}$, and $\mathrm{d}t=\nu\mathrm{d}\tilde{t}$. This gives
\begin{equation}
    S_{XX}(\omega)=\frac{1}{\nu}\tilde{S}_{XX}(\tilde{\omega}),
\end{equation}
where the PSD in natural frequency is
\begin{equation}
    \tilde{S}_{XX}(\tilde{\omega})=\lim_{\tilde{T}\rightarrow\infty}\frac{1}{\tilde{T}}\int_0^{\tilde{T}}\int_0^{\tilde{T}}\mathrm{d}\tilde{t}_1\mathrm{d}\tilde{t}_2e^{-i\tilde{\omega}(\tilde{t}_1-\tilde{t}_2)}\left\langle X(\tilde{t}_1)X(\tilde{t}_2)\right\rangle.
\end{equation}
Then $\tilde{S}_{XX}$ has units $X^2$, which will be dimensionless if $X$ is a dimensionless quantity.

With this, we can now compute the PSD of the noise $\tilde{\xi}$ using its correlation function \cref{eq:NDNoiseCorrelation}:
\begin{equation}
    \begin{aligned}
        \tilde{S}_{\tilde{\xi}\tilde{\xi}} &= \lim_{\tilde{T}\rightarrow\infty}\frac{1}{\tilde{T}}\int_0^{\tilde{T}}\int_0^{\tilde{T}}\mathrm{d}\tilde{t}_1\mathrm{d}\tilde{t_2}e^{-i\tilde{\omega}(\tilde{t}_1-\tilde{t}_2)}\left\langle\tilde{\xi}(\tilde{t}_1)\tilde{\xi}(\tilde{t}_2)\right\rangle, \\
        &=  \lim_{\tilde{T}\rightarrow\infty}\frac{1}{\tilde{T}}\int_0^{\tilde{T}}\int_0^{\tilde{T}}\mathrm{d}\tilde{t}_1\mathrm{d}\tilde{t_2}e^{-i\tilde{\omega}(\tilde{t}_1-\tilde{t}_2)}\delta(\tilde{t}_1-\tilde{t}_2), \\
        &=\lim_{\tilde{T}\rightarrow\infty}\frac{1}{\tilde{T}}\int_0^{\tilde{T}}\mathrm{d}\tilde{t}_1,\\
        &=1.
    \end{aligned}
\end{equation}
Thus white noise has a constant power spectral density:
\begin{equation}
    \tilde{S}_{\tilde{\xi}\tilde{\xi}}(\tilde{\omega})=1,
\end{equation}
representing unit power at all frequencies.

Let us now consider the position of the oscillator, which is described by the equation of motion \cref{eq:DampedOscillatorND}. We can equate the PSDs of the left- and right-hand sides. Using the constant multiple property \cref{eq:PSDConstantMultiple}, the right-hand side has PSD $\tilde{\gamma}$. To find the PSD of the left-hand side, note that it has Fourier transform
\begin{equation}
    \left[-\tilde{\omega}^2+i\tilde{\omega}\tilde{\gamma}+(2\pi)^2\right]\tilde{x}(\tilde{\omega}).
\end{equation}
Taking the absolute square we find
\begin{equation}
    \tilde{S}_{\tilde{x}\tilde{x}}(\tilde{\omega})=\frac{\tilde{\gamma}}{\left((2\pi)^2-\tilde{\omega}^2\right)^2+\left(\tilde{\omega}\tilde{\gamma}\right)^2}.
\end{equation}

To find the PSD in dimensionfull units, note that 
\begin{equation}
    S_{xx}=\ell^2S_{\tilde{x}\tilde{x}}=\frac{\ell^2}{\nu}\tilde{S}_{\tilde{x}\tilde{x}},
\end{equation}
which gives
\begin{equation}
    \begin{aligned}
        S_{xx}(\omega) &= \frac{1}{\nu}\frac{2k_BT}{m\nu^2}\frac{\tilde{\gamma}}{\left((2\pi)^2-\tilde{\omega}^2\right)^2+(\tilde{\omega}\tilde{\gamma})^2}, \\
            &= \frac{1}{\nu}\frac{2k_BT}{m\nu^2}\frac{\gamma\nu^3}{\left((2\pi\nu)^2-\omega^2\right)^2+\left(\omega\gamma\right)^2}.
    \end{aligned}
\end{equation}
Substituting $\nu=f_0$ then gives
\begin{equation}
S_{xx}(\omega) = \frac{2\gamma k_BT/m}{\left(\omega_0^2-\omega^2\right)^2+\left(\omega\gamma\right)^2}.
\label{eq:PSD fitting}
\end{equation}

\subsection{Feedback}
We can cool the system using feedback, which will be applied with some delay. The delay is partly a requirement of the electronics, which can only respond with finite speed, as well as the phase response of any FIR filter. However, feeding back a signal with the right delay can also cool the system. The feedback can be proportional to either the position, or the velocity, of the system.

Let $\Gamma_x,\Gamma_v$ denote the feedback strengths regarding to position and velocity respectively, and $\tau$ be the time delay. Then the system has the equation of motion:
\begin{equation}\label{eq:DampedOscillatorFeedback}
    \begin{aligned}
        &\ddot{x}(t)+\gamma\,\dot{x}(t)+(2\pi f_0)^2x(t)+\Gamma_xf_0x(t-\tau)+\Gamma_v\dot{x}(t-\tau) \\ &\phantom{=}=\sqrt{\frac{2\gamma k_BT}{m}}\xi(t).
    \end{aligned}
\end{equation}
The factor of $f_0$ on the position feedback term is to ensure $\Gamma_x$ has units of frequency. Note that with the signs we have chosen, $\Gamma_i$ being negative means that the position or velocity is fed back directly. For example if $\tau=0$, we could take $\Gamma_x=0$ and $\Gamma_v>0$ to cool the system using velocity feedback.

We can non-dimensionalise the system as before:
\begin{equation}\label{eq:DimensionlessFeedbackEOM}
    \tilde{x}''+\tilde{\gamma}\tilde{x}'+(2\pi)^2\tilde{x}+\tilde{\Gamma}_x\tilde{x}_{\tau}+\tilde{\Gamma}_v\tilde{x}'_{\tau}=\sqrt{\tilde{\gamma}}\tilde{\xi}.
\end{equation}
Note that we have defined
\begin{equation}
    \tilde{\Gamma}_x=\frac{\Gamma_x}{\nu},\;\tilde{\Gamma}_v=\frac{\Gamma_v}{\nu},\;\tilde{\tau}=\nu\tau,
\end{equation}
and used a subscript of $\tau$ to denote time-delay:
\begin{equation}
    \tilde{x}_{\tau}=\tilde{x}(\tilde{t}-\tilde{\tau}),\;\tilde{x}'_{\tau}=\tilde{x}'(\tilde{t}-\tilde{\tau}).
\end{equation}
To study the effect of the feedback, we must derive the power spectral density. Time delay when Fourier transformed corresponds to a phase shift:
\begin{equation}
    \begin{aligned}
        \tilde{x}_{\tau}(\tilde{t}) &\rightarrow e^{-i\tilde{\omega}\tilde{\tau}}\tilde{x}(\tilde{\omega}), \\
        \tilde{x}'_{\tau}(\tilde{t})&\rightarrow i\tilde{\omega}e^{-i\tilde{\omega}\tilde{\tau}}\tilde{x}(\tilde{\omega}).
    \end{aligned}
\end{equation}
The Fourier transform of the left-hand side thus becomes
\begin{equation}
    \left[-\tilde{\omega}^2+i\tilde{\omega}\left(\tilde{\gamma}+e^{-i\tilde{\omega}\tilde{\tau}}\tilde{\Gamma}_v\right)+\left((2\pi)^2+e^{-i\tilde{\omega}\tilde{\tau}}\tilde{\Gamma}_x\right)\right]\tilde{x}(\tilde{\omega}).
\end{equation}
The power spectral density is thus
\begin{equation}\label{eq:TimeDelayedPSD}
    \begin{aligned}
        \tilde{S}_{\tilde{x}\tilde{x}}(\tilde{\omega}) &= \tilde{\gamma}/\left(\left[(2\pi)^2-\tilde{\omega}^2+\tilde{\omega}\tilde{\Gamma}_v\sin(\tilde{\omega}\tilde{\tau})+\tilde{\Gamma}_x\cos(\tilde{\omega}\tilde{\tau})\right]^2\right. \\
        &\phantom{=}\left.+\left[\tilde{\omega}\tilde{\gamma}+\tilde{\omega}\tilde{\Gamma}_v\cos(\tilde{\omega}\tilde{\tau})-\tilde{\Gamma}_x\sin(\tilde{\omega}\tilde{\tau})\right]^2\right)
    \end{aligned} 
\end{equation}

Let us re-dimensionalise this. The overall scaling factor is
\begin{equation}
    \tilde{S}_{\tilde{x}\tilde{x}}=\frac{\nu}{\ell^2}S_{xx}.
\end{equation}
Thus we have
\begin{equation}
    \begin{aligned}
        S_{xx} &= \frac{\ell^2}{\nu}(\nu^3\gamma)/\left(\left[(2\pi\nu)^2-\omega^2+\omega\Gamma_{\nu}\sin(\omega\tau)+\nu\Gamma_x\cos(\omega\tau)\right]^2\right. \\
            &\phantom{=}\left.+\left[\omega\gamma+\omega\Gamma_v\cos(\omega\tau)-\nu\Gamma_x\sin(\omega\tau)\right]^2\right). \\
                &= (2k_BT\gamma/m)/ \\
                &\phantom{=}\left(\left[(2\pi f_0)^2-\omega^2+\omega\Gamma_v\sin(\omega\tau)+\Gamma_xf_0\cos(\omega\tau)\right]^2\right. \\
                &\phantom{=+}\left.+\left[\omega\gamma+\omega\Gamma_{\nu}\cos(\omega \tau)-f_0\Gamma_x\sin(\omega\tau)\right]^2\right).
    \end{aligned}
    \label{eq:delayed PSD fitting}
\end{equation}

\begin{figure}
\begin{center}\includegraphics[width=\linewidth]{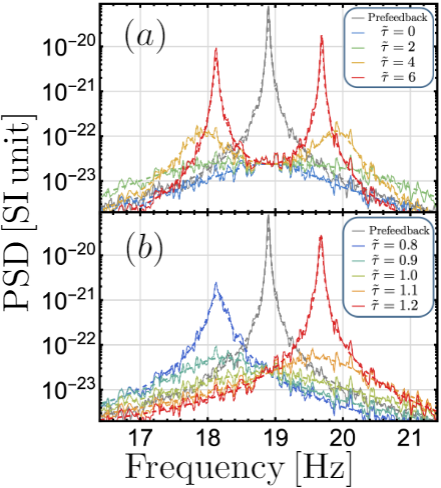}
\end{center}
\vspace*{-10pt}
\caption{\textcolor{black}{Comparison between the numerical simulations (solid lines) and analytical formula (dotted lines), \cref{eq:delayed PSD fitting}}
of the power spectral density with white Gaussian noise and delayed feedback. 
Parameters used for the simulations are: $f_0$ = 18.9 Hz, $m$ = 50 mg, $T$ = 300 K, $\tilde \gamma$ = 0.01, $\tilde \Gamma_v$ = 0.5, $\tilde \Gamma_x$ = 0.
(a) Full-period delays. As $\tilde \tau$ increases, two side peaks increase.
(b) Deviations around $\tilde \tau = 1.0$. Shifting of the peak can be observed from left of $f_0$ to right as $\tilde \tau$ increases from 0.8 to 1.2.
}
\label{fig:PSDs-Simulink-fit}
\end{figure}

Let us focus on the case of deal velocity feedback. Then in \cref{eq:TimeDelayedPSD} we have $\tilde{\Gamma}_x=0$. In the ideal case the time delay is small, and we can approximate $\sin(\tilde{\omega}\tilde{\tau})\approx\tilde{\omega}\tilde{\tau}$, and $\cos(\tilde{\omega}\tilde{\tau})\approx 1$. Then the power spectral density is
\begin{equation}
    \tilde{S}_{\tilde{x}\tilde{x}}=\frac{\tilde{\gamma}}{((2\pi)^2-\tilde{\omega}^2(1-\tilde{\Gamma}_v\tilde{\tau})^2)^2+\tilde{\omega}^2(\tilde{\gamma}+\tilde{\Gamma}_v)^2}.
\end{equation}
By minimising the denominator, we find that this is peaked at:
\begin{equation}
    \tilde{\omega}_p=\frac{\sqrt{8\pi^2(1-\tilde{\Gamma}_v\tilde{\tau})^2-(\tilde{\gamma}+\tilde{\Gamma}_v)^2}}{\sqrt{2}(1-\tilde{\Gamma}_v\tilde{\tau})^2}.
\end{equation}
Thus the damping and feedback will shift the resonant frequency. 

\textcolor{black}{We performed numerical simulations verify this analysis. We created a model of \cref{eq:DimensionlessFeedbackEOM} in SIMULINK, numerically evaluated the PSD, and compared it with our analytic form \cref{eq:TimeDelayedPSD}. The results of these are shown in \cref{fig:PSDs-Simulink-fit}. We can see that for time delays $n-1/4\lesssim \tilde{\tau}\lesssim n+1/4$ for $n$ integer we have excellent agreement between simulation and theory. However for values outside this range the simulated system blows up. This is because if the velocity is fed back with an incorrect phase, the feedback adds to the particle's velocity. This causes amplification of the system, leading to blow up. Our formula \cref{eq:TimeDelayedPSD} was derived using the Fourier transform, which assumes the system is in a steady state. In the amplification regime the assumptions of the Fourier transform are no longer valid, and hence neither is the formula. In the experiment we operate in the regime $n-1/4\lesssim\tilde{\tau}\lesssim n+1/4$, in which case simulation shows excellent agreement with our formula.}

\begin{figure*}
    \begin{center}	
    \includegraphics[width=\linewidth]{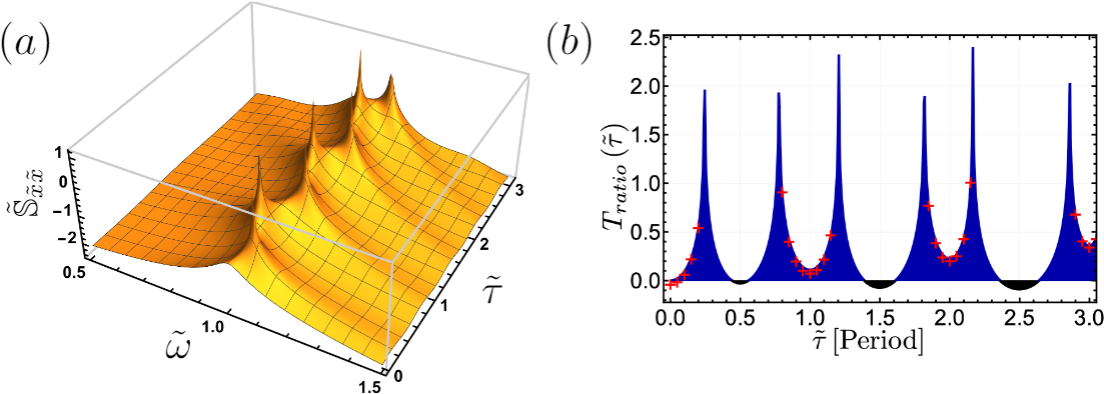}
    \end{center}
    \caption{Power spectra and temperature dependence of the oscillator with delayed feedback. (a) Plot of the {\em normalised} PSD $\tilde{\mathbb{S}}_{\tilde{x}\tilde{x}}(\tilde{\tau})$ as a function of the delay $\tilde{\tau}$. For no delay, we observe the normal Lorentzian PSD, but as we increase the delay, the system undergoes heating and cooling. (b) Performance of feedback cooling with respect to time-delay $\tilde{\tau}$. We plot $T_{ratio}(\tilde{\tau})$ defined in \cref{eq:tratio}, which quantifies the dependence of $\overline{\tilde{x}^2}$, on the delay. This vanishes when $\overline{\tilde{x}^2}(\tilde{\tau})=\overline{\tilde{x}^2}(0)$, representing cooling performance equal to the case of zero delays. The shaded curve gives the value calculated from \cref{eq:TimeDelayedPSD}, while the red crosses indicate integration from numerical simulation. The blue region indicates where $\overline{\tilde{x}^2}(\tilde{\tau})$ exceeds the value for $\tilde{\tau}=0$ where there is no delay. 
    The regions surrounding the black sections are devoid of numerical red markers. These regions are actually unstable and correspond to heating - and for these values of the delay the analytic PSD formula fails.
    The numerical data indicate the regions of stable delay periods. 
    The integration of the power spectra is done within the domain of $f_{min,max}=f_0\pm 5$ [Hz] where $f_0=18.9$ [Hz] indicates the resonant frequency.
    We choose $(\tilde{\gamma},\tilde{\Gamma}_{\nu})=(.01,0.5)$, but the behavior of this plot is sensitive to these values.}
    \label{fig:PSDs}
\end{figure*}

Next we wish to explore the dependence of the PSD on the delay time and define $\tilde{\mathbb{S}}_{\tilde{x}\tilde{x}}=\tilde{S}_{\tilde{x}\tilde{x}}\times [ (2\pi)^2 (\tilde{\gamma}+\tilde{\Gamma})/\tilde{\gamma}]/(\pi/2)$. One can show that regardless of $\tilde{\gamma},\tilde{\Gamma}$, we have $\int_0^\infty\;{\rm d}\tilde{\omega}\,\tilde{\mathbb{S}}_{\tilde{x}\tilde{x}}(\tilde{\tau}=0)=1$. This indicates that the area under the PSD for instantaneous feedback cooling is only a function of the temperature. However in general this area depends on the delay time $\tilde{\tau}$. To investigate this we plot in \cref{fig:PSDs}a) the {\em normalised}, PSD $\tilde{\mathbb{S}}_{\tilde{x}\tilde{x}}$, to see how it changes with $\tilde{\tau}$, for specific values of $(\tilde{\gamma}, \tilde{\Gamma})$. We recall that the integral of the PSD quantifies the position variance $\overline{x^2}\approx \int\,d\omega\, S_{xx}(\omega,\tau)$. We can define
\begin{equation}
T_{ratio}(\tilde{\tau})\equiv \log_{10}\left[\int_0^\infty\;{\rm d}\tilde{\omega}\,\tilde{\mathbb{S}}_{\tilde{x}\tilde{x}}(\tilde{\tau}) \right]\label{eq:tratio}
\end{equation}
which logarithmically quantifies the ratio of the areas under the feedback-PSD with delay to that with no delay. We interpret this as the temperature of the oscillator. We plot $T_{ratio}$ when the parameters have the example values $(\tilde{\gamma},\tilde{\Gamma}_{\nu})=(.01,0.5)$, but we note that the details of this plot are highly dependent on these values. We generally note that there are delay values for which $\overline{x^2}$ essentially diverges while there are other delay values where $\overline{x^2}$ is slightly lower than when $\tilde{\tau}=0$. In \cref{fig:PSDs}(b) we plot $T_{ratio}(\tilde{\tau})$, using the analytic form of the delayed-PSD and note that there are some regions of delays where the oscillator is cooled below that with instantaneous feedback and there are delay values where the oscillator is heated. 
We define the effective temperature of the particle's motion for the experiment as the ratio of the areas under the PSD e.g.
\begin{equation}
T_{eff}=300\times \frac{\tilde{\gamma}}{\tilde{\gamma}+\tilde{\Gamma}_v}\times T_{ratio}(\tilde{\tau})\;\;{\rm [{}^\circ K]}
\label{eq:temperature_eq}
\end{equation}.

\section{Measurements of PSD}\label{Section:PSD_measurement}
The position of the resonator is measured at a sampling frequency of $2.44 \:\rm kHz$ using the PicoScale interferometer, which is mostly sensitive to vertical motion and we focus on the vertical mode of motion of the resonator. The PSD is estimated using Welch's method. 
The resonators have Q factors above $10^5$ in high vacuum with a decay time of $ \sim 3000 \:\rm s$. For time efficiency, we first studied the delayed feedback cooling at moderate pressure ($\sim 1\times10^{-2} \:\rm mBar$) and collected data around $20 \:\rm minutes$ for each measurement. Then we moved to high vacuum ($\sim 1\times10^{-6} \:\rm mBar$) and redid the measurement but with a longer data collection time of $\sim7 \:\rm hours$ to have a better resolution of the high-Q PSD. \textcolor{black}{We observed that the PSD vertical mode peak drifts towards slightly higher frequencies during the $7 \:\rm hours$, leading to a peak width expansion if one uses the entire acquisition time to compute the PSD. As shown in Fig.\ref{fig:PSD_1hSection}, the PSD peak of each $1 \:\rm hour$ chunk drifts as time goes on and contributes to a broad peak for the PSD when computed using the whole $7 \:\rm hours$ data. The cause of this drift is not clear yet, but we suspect it may be caused by the thermal expansion effects related either to the composite resonator or the surrounding apparatus in the high vacuum.}

\section{Analysis of PSD and Fit to Theory}\label{Section:PSD_theory_analysis}

\subsection{Estimation of delay introduced from the FIR filter}
%%%%%%%%%%%%<<<FIGURE>>>%%%%%%%%%%%%%%%%%
\begin{figure}[h!]
\begin{center}	\includegraphics[width=\linewidth]{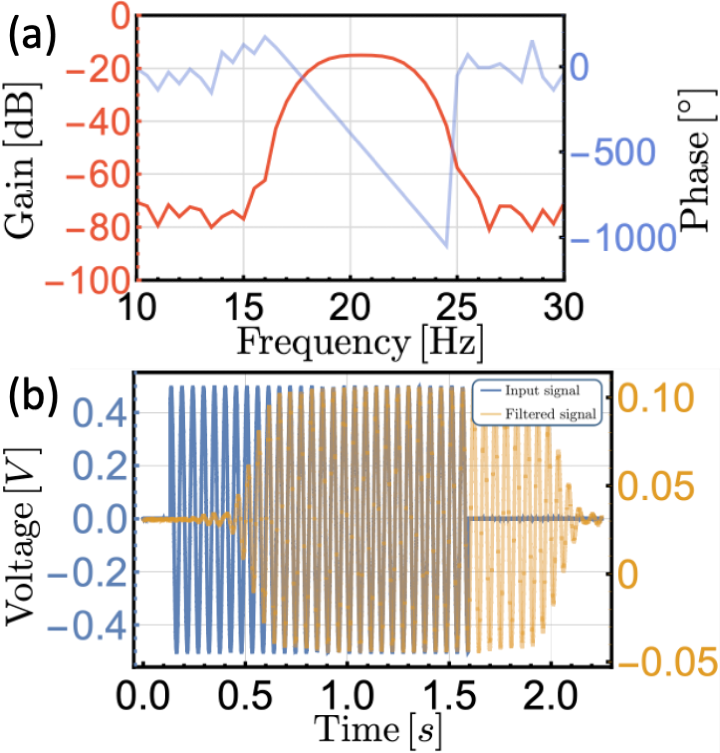}
\end{center}
\vspace*{-10pt}
\caption{Characterization of the single-rate finite impulse response (FIR) bandpass filter employed in feedback cooling.
(a) Frequency response of the bandpass filter. The plot demonstrates the amplitude and phase characteristics of the filter, highlighting its targeted performance at approximately 18.9 Hz, corresponding to the vertical mode. The filter effectively attenuates frequencies outside the passband, mitigating potential reductions in cooling efficiency.
(b) Signal response of the bandpass filter for an 18.9 Hz sine signal, showing an intrinsic $\sim 7.6$ periods delay for filtered signal. In the experiment, a manual fractional delay is added to match the phase of the input and output signals. The signal attenuation results from two sources: the filter's amplitude response and a manual adjustment made to facilitate binary calculations in the FPGA. A manual DC shift is also applied to the filtered signal to optimize the magnetic actuation.}
    \label{fig:Filter_Signal_Response}
\end{figure}

%%%%%%%%%%%%<<<FIGURE>>>%%%%%%%%%%%%%%%%%
\begin{figure}[h!]
\begin{center}	\includegraphics[width=\linewidth]{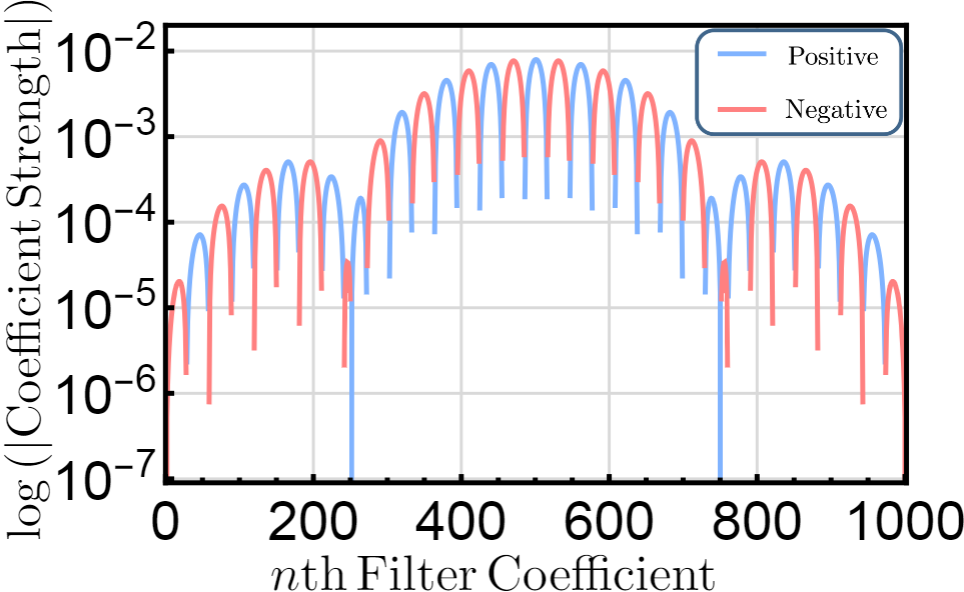}
\end{center}
\vspace*{-10pt}
\caption{Filter coefficients used in the bandpass filter. The blue (red) plot indicates the positive (negative) sign of the coefficient strength. The coefficient strength shows the characteristic symmetry of an FIR filter and is strongest in the middle.}
    \label{fig:MikeFilterCoeff}
\end{figure}
%%%%%%%%%%%%%%%%%%%%%%%%%%%%%%%%%%%%%%%%%

The single-rate FIR filter operates similarly to a weighted moving average determined by its filter coefficients. 
As a result, the influence of signal data extends over the length of these coefficients, and the filter's sampling frequency determines the time duration in seconds.
This process introduces a delay in the output signal, as depicted in Fig. \ref{fig:Filter_Signal_Response}.
The delayed signal resulting from the filter is distinct from a manually delayed signal.
However, because our filter's coefficients are symmetric, with the most significant contribution occurring in the middle, as depicted in Fig. \ref{fig:MikeFilterCoeff}, we can approximate the filtered signal using the delayed equation where the delay is half the total time the signal influences the output.

With a filter sampling frequency of 1250 Hz (equivalent to 1250 samples per second) and a filter length of 1001 coefficients, and assuming a resonant frequency of 19 Hz (19 periods per second), we can calculate half the duration over which a signal will influence the output in terms of the number of periods $\tilde{\tau}_{est}$.
\begin{equation}
    \tilde{\tau}_{est}= \frac{1}{2}* 1001\, [\text{samples}]* \frac{19\, [\text{periods/s}]}{1250\, [\text{samples/s}]} \approx 7.6 \text{ periods}.
\end{equation}
Since we introduced an additional manual delay for phase matching, as detailed in the experimental methods section, we estimated the filtered signal using a delayed analytical equation with an 8-period delay unless manually adjusted.

\subsection{PSD of experimental data and fitting}

We measure the PSD at moderate and low pressures with different time delays and feedback strengths. The experiment data matches the theory with good fitting to \cref{eq:delayed PSD fitting}. In \cref{eq:delayed PSD fitting} there are five parameters, the scaling factor $S=2k_{B}T/m$, the natural frequency $f_{0}$, the damping frequency $\gamma$, the feedback strength $\Gamma_{v}$, and the time delay $\tau$. To reduce the number of degrees of freedom, and hence have more confidence in the fitting, we first fix some of the known parameters. From the fitting of the feedback-off data to \cref{eq:PSD fitting}, we estimate the damping rate $\gamma$ and we fix it for all of the fitting of feedback-on data to \cref{eq:delayed PSD fitting}. For the series of data comparing different time delays, we restrict the feedback strength to a minimal range, since we keep the feedback strength consistent in the experiment. For the series of data comparing different feedback strengths, we strictly restrict the time delay to be around 8 periods in the fitting, as we set 8 periods delay for these measurements. 

\begin{figure}[h!]
\begin{center}	\includegraphics[width=\linewidth]{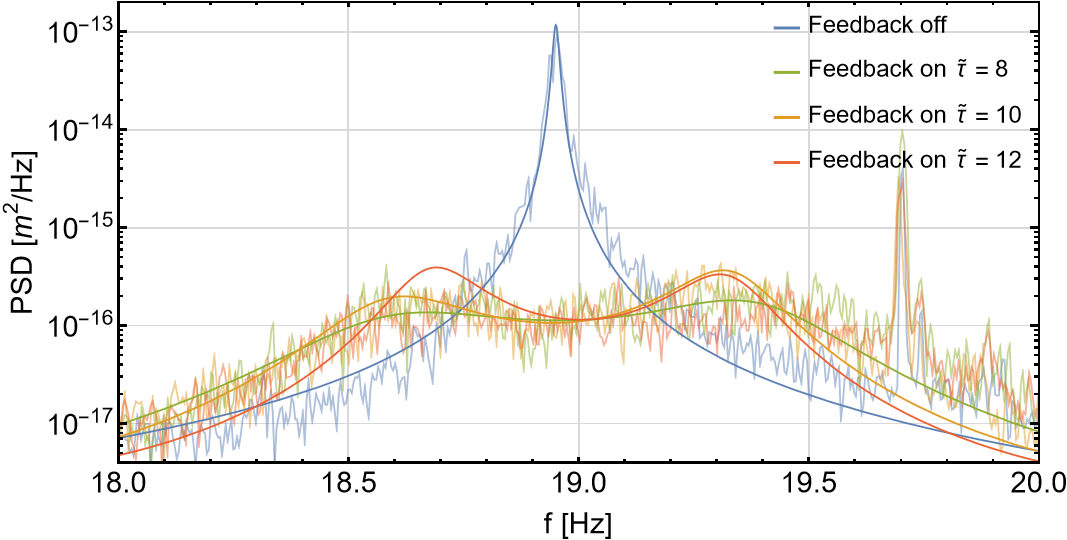}
\end{center}
\vspace*{-10pt}
\caption{PSD of the vertical mode at moderate pressure ($\sim 1\times 10^{-2} \:{\rm mBar}$) with different integer delays ranging from $\tilde \tau=8,\: 10,\: 12$.
The noisy curves denote the experimental data, and the smooth curves show fittings to the theory. }
    \label{fig:HP_delay_integer}
\end{figure}

\begin{figure}[h!]
\begin{center}	\includegraphics[width=\linewidth]{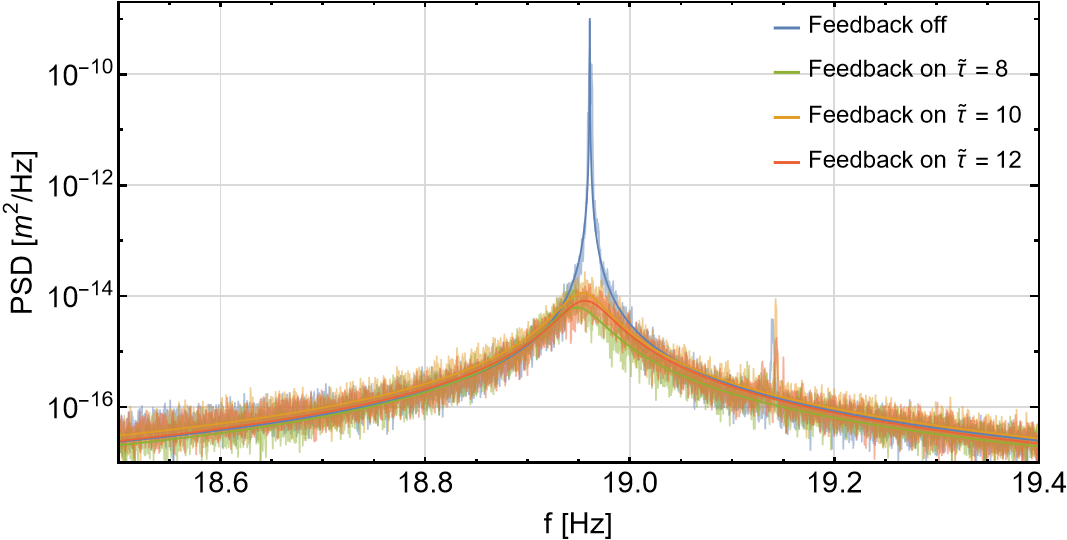}
\end{center}
\vspace*{-10pt}
\caption{PSD of the vertical mode at low pressure ($\sim 1\times 10^{-6} \:{\rm mBar}$) with different integer delays ranging from $\tilde \tau=8,\: 10,\: 12$.
The noisy curves denote the experimental data, and the smooth curves show fittings to the theory.} 
    \label{fig:LP_delay_integer}
\end{figure}

The experiment data is in good agreement with the theory for delayed feedback cooling. As there is an intrinsic $\sim 7.6$ period delay for a $19\: {\rm Hz}$ oscillator in the circuit, we choose to add an additional delay and the smallest integer delay we can get is thus $8$ periods. At moderate pressure, we fix the feedback strength at $\Gamma_v=3\:{\rm Hz}$ and set the delay to be $\tau/T=7.9,\: 8,\: 8.1$ respectively, where $T$ is the mechanical period. We find that the strongest cooling is achieved when the delay $\tau/T$ is an integer, and we show these PSDs at moderate and low pressures in \cref{fig:HP_delay_integer} and \cref{fig:LP_delay_integer}. The mode peaks tend to get broader, and side peaks will be driven when $\tau/T$ deviates from an integer value. The motions tends will also get more heated with increasing integer delay periods.  Our experimental observations confirm the theoretical modeling shown in \cref{fig:PSDs}.

\begin{figure}[h!]
\begin{center}	\includegraphics[width=\linewidth]{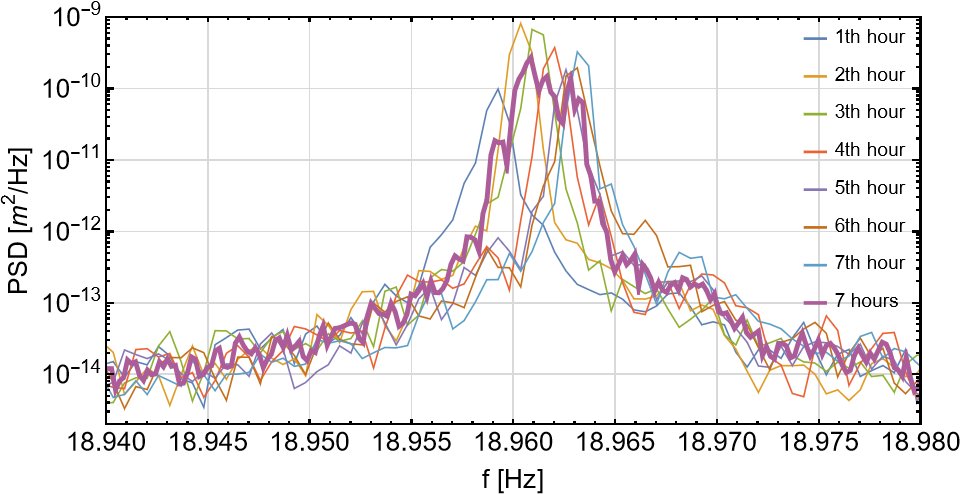}
\end{center}
\vspace*{-10pt}
\caption{Peak of the vertical mode PSD at low pressure ($\sim 1\times 10^{-6} \:{\rm mBar}$) for each 1 hour chunk. The PSD peak of each $1 \:\rm hour$ chunk drifts to high frequency as time goes on and contributes to a broad peak for the PSD of the whole $7 \:\rm hours$ data. From this we deduce that there are slow drifts in the trapping forces - whose origin are as yet unknown.}
    \label{fig:PSD_1hSection}
\end{figure}

With the increase of the feedback strength, the PSD peak of the vertical mode decreases. However, side modes will be driven due to the delay in the feedback signal, if the feedback strength is too strong. 
The effects of delay on the cooling are the same for the measurements at low pressure. 
We consider the undriven resonator to be in thermal equilibrium with the environment ($300 \: {\rm K}$) and by comparing the areas under the the fitted PSDs (see main text Fig. 4d), we estimate the resonator to be cooled to $\sim 320\: {\rm mK}$ with appropriate delay and feedback strength. Finally, we acquired data over 7 hours and divided this into separate 1 hour chunks. We analysed each chunk to obtain the PSD for each hour of acquisition and we superimpose these plots together in \cref{fig:PSD_1hSection}. From this we notice a slow drift in the trap frequency over time whose origin is as yet undetermined but most likely due to thermal effects and nonlinearities in the system.

\begin{table*}[!hbt]
    \centering
    \caption{Fitting parameters for PSD measured at low pressure ($\sim 10^{-6} \:\rm mBar$)}
    \begin{tabular}
    {p{2cm}p{1.2cm}p{1.2cm}p{1.1cm}p{1.2cm}p{0.9cm}p{1.2cm}p{1.4cm}p{1.2cm}p{0.75cm}p{0.9cm}p{1.2cm}p{1.2cm}p{1.2cm}}
    \hline\hline
        Parameters & \multicolumn {2}{c}{Scaling factor} & \multicolumn {2}{c}{Damping frequency}  &\multicolumn {2}{c}{ Natural frequency} & \multicolumn {3}{c}{Delay time} & \multicolumn {2}{c}{Feedback strength}  & \multicolumn {2}{c}{Area under PSD}  \\ \hline
        Symbols & S & error & $\gamma (Hz)$ & error  & $f_{0} (Hz)$ & error & $\tau (s)$ & error & $\tilde\tau$ & $\Gamma_{v} (Hz)$ & error & A & error \\ \hline
        Feedback off & 1.57E-08 & 6.92E-33 & 7.0E-04 & 5.41E-06 & 18.961 & 2.03E-05 & $\backslash$ & $\backslash$ & $\backslash$ & $\backslash$ & $\backslash$ & 2.76E-13 & 8.61E-23 \\ 
        F/B on, $\Gamma_{v}=0.3Hz, T=8$ & 1.17E-08 & 6.48E-11 & 7.0E-04 & 5.41E-06 & 18.951 & 3.24E-03 & 4.214E-01 & 1.09E-03 & 7.99 & 0.304 & 5.60E-03 & 4.89E-16 & 1.11E-25 \\ 
        F/B on, $\Gamma_{v}=0.3Hz, T=7.9$ & 1.78E-08 & 1.18E-10 & 7.0E-04 & 5.41E-06 & 18.952 & 4.02E-03 & 4.170E-01 & 1.10E-03 & 7.91 & 0.301 & 3.04E-02 & 9.38E-16 & 1.10E-26 \\ 
        F/B on, $\Gamma_{v}=0.3Hz, T=8.1$ & 2.11E-08 & 2.42E-10 & 7.0E-04 & 5.41E-06 & 18.949 & 6.42E-03 & 4.272E-01 & 1.74E-03 & 8.10 & 0.307 & 4.72E-02 & 1.07E-15 & 5.85E-27 \\ 
        F/B on, $\Gamma_{v}=0.3Hz, T=10$ & 1.69E-08 & 1.02E-10 & 7.0E-04 & 5.41E-06 & 18.941 & 2.25E-03 & 5.318E-01 & 6.70E-04 & 10.08 & 0.307 & 1.38E-02 & 8.04E-16 & 2.56E-24 \\ 
        F/B on, $\Gamma_{v}=0.3Hz, T=12$ & 1.35E-08 & 7.86E-11 & 7.0E-04 & 5.41E-06 & 18.946 & 1.57E-03 & 6.364E-01 & 4.91E-04 & 12.07 & 0.307 & 8.30E-03 & 6.24E-16 & 2.22E-24 \\
        F/B on, $\Gamma_{v}=0.15Hz, T=8$ & 1.68E-08 & 1.16E-10 & 7.0E-04 & 5.41E-06 & 18.952 & 4.00E-03 & 4.224E-01 & 3.17E-03 & 8.01 & 0.132 & 3.89E-03 & 1.59E-15 & 3.01E-25 \\ 
        F/B on, $\Gamma_{v}=0.75Hz, T=8$ & 1.61E-08 & 2.96E-10 & 7.0E-04 & 5.41E-06 & 18.946 & 1.06E-02 & 4.224E-01 & 1.45E-03 & 8.01 & 0.725 & 2.08E-02 & 2.94E-16 & 1.11E-24 \\ 
        F/B on, $\Gamma_{v}=1.5Hz, T=8$ & 4.50E-08 & 1.51E-09 & 7.0E-04 & 5.41E-06 & 18.984 & 1.84E-02 & 4.214E-01 & 9.31E-04 & 7.99 & 1.760 & 5.56E-02 & 3.76E-16 & 4.48E-30 \\
        \hline
    \end{tabular}
    \label{LP_fitting}
\end{table*}

\begin{table*}[!ht]
    \centering
    \caption{Fitting parameters for PSD measured at moderate pressure ($\sim 10^{-2}\:\rm mBar$)}
    \begin{tabular}
{p{2cm}p{1.2cm}p{1.2cm}p{1.1cm}p{1.2cm}p{0.9cm}p{1.2cm}p{1.4cm}p{1.2cm}p{0.75cm}p{0.9cm}p{1.2cm}p{1.2cm}p{1.2cm}}
    \hline\hline
        Parameters & \multicolumn {2}{c}{Scaling factor} & \multicolumn {2}{c}{Damping frequency}  &\multicolumn {2}{c}{ Natural frequency} & \multicolumn {3}{c}{Delay time} & \multicolumn {2}{c}{Feedback strength}  & \multicolumn {2}{c}{Area under PSD}  \\ \hline
        Symbols & S & error & $\gamma (Hz)$ & error  & $f_{0} (Hz)$ & error & $\tau (s)$ & error & $\tilde\tau$& $\Gamma_{v} (Hz)$ & error & A & error \\ \hline
    
        Feedback off & 1.50E-10 & 2.16E-11 & 9.1E-02 & 1.36E-02 & 18.951 & 1.12E-03 & $\backslash$ & $\backslash$ & $\backslash$ & $\backslash$ & $\backslash$ & 2.64E-15 & 9.23E-24 \\ 
        F/B on, $\Gamma_{v}=3Hz, T=8$ & 1.83E-10 & 4.76E-12 & 9.1E-02 & 1.36E-02 & 19.001 & 8.64E-03 & 4.216E-01 & 2.42E-04 & 7.99 & 3.102 & 5.71E-02 & 1.75E-16 & 2.29E-27 \\ 
        F/B on, $\Gamma_{v}=3Hz, T=7.9$ & 2.71E-10 & 1.14E-11 & 9.1E-02 & 1.36E-02 & 18.982 & 1.17E-02 & 4.166E-01 & 2.64E-04 & 7.90 & 3.105 & 1.28E-01 & 4.20E-16 & 4.88E-24 \\ 
        F/B on, $\Gamma_{v}=3Hz, T=8.1$ & 1.54E-10 & 4.06E-12 & 9.1E-02 & 1.36E-02 & 18.988 & 1.42E-02 & 4.259E-01 & 2.01E-04 & 8.07 & 3.099 & 1.38E-01 & 2.32E-16 & 5.40E-27 \\ 
        F/B on, $\Gamma_{v}=3Hz, T=10$ & 1.72E-10 & 3.55E-12 & 9.1E-02 & 1.36E-02 & 18.980 & 4.79E-03 & 5.278E-01 & 1.51E-04 & 10.00 & 3.099 & 4.47E-02 & 2.11E-16 & 6.06E-25 \\ 
        F/B on, $\Gamma_{v}=3Hz, T=12$ & 1.50E-10 & 4.40E-12 & 9.1E-02 & 1.36E-02 & 18.994 & 5.72E-03 & 6.316E-01 & 2.09E-04 & 11.97 & 2.792 & 6.19E-02 & 2.19E-16 & 1.42E-24 \\ 
        F/B on, $\Gamma_{v}=0.75Hz, T=8$ & 1.97E-10 & 5.50E-12 & 9.1E-02 & 1.36E-02 & 18.972 & 1.50E-02 & 4.216E-01 & 1.95E-03 & 7.99 & 0.764 & 3.56E-02 & 4.21E-16 & 4.40E-24 \\ 
        F/B on, $\Gamma_{v}=1.5Hz, T=8$ & 1.64E-10 & 4.33E-12 & 9.1E-02 & 1.36E-02 & 18.988 & 1.21E-02 & 4.216E-01 & 7.75E-04 & 7.99 & 1.492 & 4.46E-02 & 2.18E-16 & 2.46E-27 \\ 
        F/B on, $\Gamma_{v}=6Hz, T=8$ & 1.68E-10 & 3.43E-12 & 9.1E-02 & 1.36E-02 & 19.018 & 3.29E-03 & 4.211E-01 & 4.65E-05 & 7.98 & 6.498 & 3.59E-02 & 4.12E-16 & 1.99E-24 \\  
    \hline
    \end{tabular}
    \label{HP}
\end{table*}

\begin{table*}[hbt!]
\begin{center}
\caption{Parameters pertaining to the novel insulating resonator used in the works of this paper\label{table:resonator}}
\begin{threeparttable}
\begin{tabular}{p{3cm}p{3cm}p{2cm}}%p{3cm}}
\hline\hline
  Quantity & Value & Unit   \\ \hline
 Thickness & (0.530 $\pm$ 0.008) & mm \\
 Length & (7.88 $\pm$ 0.05) & mm \\ 
 Width & (7.97 $\pm$ 0.01) & mm \\
 Mass & 49 & mg \\
 Conductivity & < 8.6 x $10^{-9}$ & S/m \\
\hline
\end{tabular}

\end{threeparttable}
\end{center}
\vspace{-7mm}
\end{table*}

\section*{References}
 %\bibliography{references2}
%aipnum4-2.bst 2019-01-14 (MD) hand-edited version of apsrev4-1.bst